\documentclass[journal,transmag]{IEEEtran}
\ifCLASSINFOpdf
\else
\fi
\pdfoutput=1

\usepackage{amsmath,amssymb,bm,graphicx,float,color,subfigure}
\usepackage[normalem]{ulem}

\begin{document}
%
\title{Importance sampling for thermally induced switching and
  non-switching probabilities in spin-torque magnetic nanodevices}


\author{\IEEEauthorblockN{ YiMing Yu, Cyrill B. Muratov, and Richard
    O. Moore} \IEEEauthorblockA{Department of Mathematical Sciences,
    New Jersey Institute of Technology, Newark, NJ 07102,
    USA}
}

%



\IEEEtitleabstractindextext{%
\begin{abstract}
  Spin-transfer torque magnetoresistive random access memory is a
  potentially transformative technology in the non-volatile memory
  market. Its viability depends, in part, on one's ability to
  predictably induce or prevent switching; however, thermal
  fluctuations cause small but important errors in both the writing
  and reading processes.  Computing these very small probabilities for
  magnetic nanodevices using naive Monte Carlo simulations is
  essentially impossible due to their slow statistical convergence,
  but variance reduction techniques can offer an effective way to
  improve their efficiency.  Here, we provide an illustration of
    how importance sampling can be efficiently used to estimate
  low read and write soft error rates of macrospin and
  coupled-spin systems.
\end{abstract}

}

\maketitle

\IEEEdisplaynontitleabstractindextext

%
\IEEEpeerreviewmaketitle

\section{Introduction}
%
%
%
%
\IEEEPARstart{S}{pin-transfer} torque magnetoresistive random access
memory (STT-MRAM) has been proposed as a non-volatile replacement for
random access memory that offers high speed, low power consumption,
non-volatility and unlimited endurance
\cite{zhu08,Khvalkovskiy2013,kent15,apalkov16,bhatti17}.  One of the
primary obstacles to its widespread deployment is physical scaling,
due to an increased error rate that accompanies smaller volumes of
storage cells.  A memory device should switch quickly and reliably
when switching is intended and otherwise maintain its current
state. However, thermal fluctuations in the magnetization orientation
can sometimes induce unwanted switching during either storage or an
attempted read event, or failure to switch during an attempted write
event. These contribute to the write soft error rate (WSER), read soft
error rate (RSER), and retention failure rate \cite{Butler2012}.  The
expected values of WSER and RSER in STT-MRAM should not exceed the
order of $10^{-18}$ without error correction~\cite{Khvalkovskiy2013}.
Due to the importance of these extremely small rates in quantifying
the viability of experimental STT-MRAM configurations, analytical and
computational techniques that facilitate their calculation are
critically important.

One approach makes use of the Fokker-Planck equation (FPE) describing
the time evolution of the switching probability
\cite{Xie2017,Tzoufras2018,XieMa2017}.  In the macrospin approximation,
treating each STT-MRAM bit as a single magnetic domain, the FPE can be
solved directly or can be further approximated by the Brown-Kramers
formula, which overestimates the RSER for short read
times~\cite{Butler2012}. However, both the effects of spatial
variations in the magnetization across a single memory cell and
interactions between adjacent cells obviously cannot be captured in
the macrospin approximation. These effects increase in importance as
the size of each cell exceeds the scale of above $50$ nm in lateral
size and must therefore be taken into account to support development
of magnetic nanodevices at this scale \cite{Sun2011}.  Direct
numerical simulations of the FPE increase exponentially in
computational cost as the dimension of the coupled system of
macrospins is increased, and this high cost is exacerbated by the
presence of boundary layers associated with the small size of thermal
fluctuations.  Efficient computational methods are therefore needed to
provide a means of determining these small switching probabilities and
rates.

The most common approach to computing switching probabilities uses
sampling to provide an empirical estimate of the quantity of
interest. However, the extremely low switching probabilities and rates
relevant to micromagnetic devices make naive Monte Carlo studies
essentially impossible. A common approach to recover the tails of the
probability distribution from Monte Carlo simulations is via
extrapolation (see, e.g., \cite{heindl11, Zhao2012,
  shiota16}). However, this may introduce large uncontrolled
inaccuracies due to failure of the fitting form to capture the
asymptotic behavior of the probability distribution in the small noise
limit \cite{marcuse_derivation_1990,moore_method_2008}. Alternatively,
variance reduction techniques such as importance splitting attempt to
concentrate the samples generated on those with a higher likelihood of
registering a rare event of interest \cite{landau_guide_2000, dean09,
  rubinstein_simulation_2017, Roy2016,PRAMANIK201896}.

Here, we demonstrate that accurate switching probabilities and error
rates of STT-MRAM devices can be computed efficiently using another
variance reduction technique known as importance sampling (IS) in an
ensemble of biased Monte Carlo simulations. We intentionally choose a
particularly simple micromagnetic setting to clearly illustrate the IS
methodology in the context of STT-MRAM modeling. The main point of
this paper is to demonstrate that with the help of IS the events with
extremely low probability of occurrence can be accessed via direct
Monte Carlo sampling with an additional post-processing step. This,
together with the simplicity of its implementation could make the IS
based approach a powerful tool in assisting the design of the next
generation of spintronic nanodevices.

Our paper is organized as follows. In Section \ref{s:IS}, we review IS
and the strategies for optimal choices of bias. In Section
\ref{s:micro} we formulate the stochastic micromagnetic model for a
single macrospin or a system of exchange coupled macrospins, and then
show how to apply IS to sample switching probabilities in the
macrospin approximation. In Section \ref{s:simulations}, we carry out
IS simulations for a single macrospin and a coupled system of two
indentical macrospins with different choices of biases to demonstrate
the power and efficiency of IS in the context of STT-MRAM
applications. Finally, in Section \ref{s:discussion} we briefly
summarize our findings.

\section{Importance sampling} \label{s:IS}

Suppose we wish to estimate the probability
$P={\mathbb{E}_0}[I(\omega)]$ of a system driven by random variable
$\omega$ producing an event with indicator function $I(\omega)$.
Here, ${\mathbb{E}}_0[\cdot]$ denotes the expected value with respect
to the density $\rho_0(\omega)$, such that
\begin{equation}
P = {\mathbb{E}}_0[I(\omega)]=\int I(\omega)\rho_0(\omega)\,d\omega.
\end{equation}
A naive Monte Carlo method uses $M$ independent draws $\omega_i$ from
$\rho_0$ to approximate $P$ according to
\begin{equation}
  P_{MC} = \frac 1 M \sum_{i=1}^M I(\omega_i).
\end{equation}
The coefficient of variation (CV), which measures the statistical
error of the sample, is given by
\begin{equation}
  CV(P_{MC}) = \frac{\sqrt{Var(P_{MC} )}}{\mathbb{E}_0[P_{MC}]}  =
  \frac{1}{\sqrt{M}} \sqrt{\frac{1}{ P_{MC}}-1}.
\end{equation}
For $P_{MC}\ll 1$, smallness of $CV(P_{MC})$ requires
$M\gg 1/P_{MC} \gg 1$.  In the present case, this necessitates an
impractically large number of simulations to produce a reasonable
estimate of the probability.  The idea behind IS to sample $\omega$
from an alternative probability density $\rho_u (\omega)$ that depends
on a bias $u$ chosen to increase the likelihood of the event of
interest. An unbiased estimator is then recovered by weighting each
result according to
\begin{equation}
  P_{IS} = \frac 1 M \sum_{i=1}^M I(\omega_i)L(\omega_i),
 \label{e:ISestimator}
\end{equation}
where $L(\omega) = {\rho_0(\omega)}/{\rho_u(\omega)}$ is called the
likelihood ratio, and assumes that $\rho_u(\omega) >0$ whenever
$\rho_0(\omega) > 0.$ The resulting CV is then given by
\begin{equation}
  CV(P_{IS}) = \frac{\sqrt{Var(P_{IS} )}}{\mathbb{E}_{u}[P_{IS}]}  =
  \frac{1}{\sqrt{M}}\sqrt{\frac{\mathbb{E}_{u}[I(\omega)L^2(\omega)]}{
      \mathbb{E}_{u}[P_{IS}]^2}-1},
\label{e:ISCV}
\end{equation}
where $\mathbb E_u$ denotes the expectation with respect to $\rho_u$
and $1/P \ge {\mathbb{E}_u[I(\omega)L^2(\omega)]}/P^2 \ge 1$.
Expression~(\ref{e:ISCV}) suggests that a ``good'' bias to use in IS
keeps ${\mathbb{E}_u[I(\omega)L^2(\omega)]}/P^2 $ close to 1.

In many cases of interest, including the one discussed here, the
underlying physical model is described by a stochastic differential
equation (SDE)
\begin{equation} \label{eqn:dp} dX(t) = b(X(t))dt + \epsilon
  \sigma(X(t))dW(t), \quad X(0) = X_0,
\end{equation}
where $X(t) \in \mathbb{R}^N$, for some $N\geq 1$, is a randomly
evolving state variable, $b(x)$ is its deterministic drift,
$\sigma(x)$ is the noise strength matrix, $dW$ is an infinitesimal
increment of an $N$-dimensional Brownian motion, and $\epsilon > 0$ is
the noise strength. In many situations of interest the noise is weak:
$\epsilon\ll 1$. If we wished to observe the state $X(t)$ exhibiting a
behavior that is very rare for typical realizations of the Brownian
motion $W(t)$, we would have to simulate this equation a very large
number of times, even more so to produce an accurate probability
estimate from these empirical observations.
In the case of SDEs, the IS technique makes rare events happen more
often by introducing a bias $\epsilon^{-1} u(X(t), t) dt$ to the mean
of the noise increment $dW$, such that the coefficient of variation of
the estimate $P_{IS}$ is reduced at the same time
\cite{Dupuis1987,Rubino2009}. This provides new paths $\tilde{X}(t)$
that evolve according to
\begin{align} \label{eqn:controlsystem} d \tilde{X} = \left(
    b(\tilde{X}) + \sigma(\tilde{X}) u(\tilde{X}, t)\right) dt +
  \epsilon \sigma(\tilde{X}) dW.
\end{align}
By Girsanov's theorem~\cite{cottrell_large_1983,Eric2012}, for a time
horizon $T > 0$ the likelihood ratio is given by
\begin{equation}
  \label{eq:L}
  L = \exp \left(
    -\frac{1}{2\epsilon^2}\int_0^T|u(\tilde{X}, t)|^2 dt -
    \frac{1}{\epsilon}\int_0^T \langle u(\tilde{X}, t),dW(t)
    \rangle
  \right),
\end{equation}
where $\langle \cdot, \cdot \rangle$ stands for the Euclidean inner
product in $\mathbb R^N$, $W(t)$ is the realization of the noise that
produced $\tilde X(t)$, and the last integral is understood in the
It\^o sense. This expression may then be incorporated into the IS
estimator~(\ref{e:ISestimator}) to recover an unbiased probability
estimate. We emphasize that in the limit $M \to \infty$ one recovers
from~(\ref{e:ISestimator}) the exact value of $P$ for the original,
unbiased process. The effect of good biasing is simply to concentrate
the runs sampled in~(\ref{e:ISestimator}) on those with highest
likelihood of activating the indicator function, thereby producing an
accurate estimate of $P$ with vastly fewer biased Monte Carlo runs.

Over finite-time horizons, an effective bias $u = u^*$ can be obtained
by minimizing the Freidlin-Wentzell large deviation action. Namely,
for a given time horizon $T > 0$, current time $t < T$, current state
$x$ and the set of targeted outcomes $\mathcal A$ one looks for the
minimizer $\phi_{t,x}^T(s)$ of the functional
\begin{equation}
  \label{min_Action} S_T[\phi] = \int_t^T \frac{1}{2}|
  \sigma ^{-1} (\phi(s))(\dot{\phi}(s) - b(\phi(s)))|^2 ds,
\end{equation}
where $\dot \phi(s) = d \phi(s)/ds$, over absolutely continuous paths
$\phi : [t, T] \to \mathbb R^N$ satisfying $\phi(t) = x$ and
$\phi(T) \in \mathcal A$ \cite{freidlin_random_2012}. The finite-time
bias function $u^* = u^*_T$ is then given by
\cite{Dupuis1987,Eric2012,dupuis_importance_2012}
\begin{equation}
  \label{control:1}
  u^*_T(x, t) = \sigma^{-1}
  (x)(\dot{\phi}_{t,x}^T(t) - b(x)).
\end{equation}
Over infinite-time horizons, i.e., when $T \to \infty$, a convenient
reparametrization allows the action in~(\ref{min_Action}) to be
minimized with respect to arclength rather than time. In this case,
one can choose $u^* = u^*_\infty$, where $u^*_\infty(x)$ is obtained
from the minimizer $\phi_x(\alpha)$ of the
functional~\cite{freidlin_random_2012},\cite{heymann_geometric_2008}
  \begin{align}
    \label{eq:Sinf}
    S_\infty[\phi] = \int_0^1 \big(|\sigma^{-1}(\phi(\alpha))
    \phi'(\alpha) | \, |\sigma^{-1}(\phi(\alpha)) b(\phi(\alpha))|
    \notag \\
    -
    \langle \sigma^{-1} (\phi(\alpha)) \phi'(\alpha),\sigma^{-1}
    (\phi(\alpha)) b(\phi(\alpha)) \rangle \big) 
    d\alpha, 
 \end{align}
 among all absolutely continuous paths $\phi: [0,1] \to \mathbb R^N$
 satisfying $\phi(0) = x$ and $\phi(1) \in \mathcal A$. Here
 $\phi'(\alpha) = d \phi(\alpha) / d \alpha$. The infinite-time bias
 is then given by \cite{Dupuis1987}
 \begin{align}
   \label{uinf}
   u^*_\infty(x) = \sigma^{-1}(x) \left( {|\sigma^{-1}(x) b(x)|
   \over 
   |\sigma^{-1}(x) \phi'_x(0)|} \phi'_x(0) - b(x) \right).
 \end{align}
 
One of the strategies discussed below is the use of IS with
infinite-time bias functions to obtain switching probabilities over
finite time horizons.  This strategy is based on the observation that,
when the characteristic speed obtained by dividing the domain radius
by the time horizon is small relative to the maximum speed of the
infinite-time minimizing path, the finite-time and infinite-time
minimizing paths are nearly identical outside of small neighborhoods
around the dynamic fixed points.  As will be seen in
Sec.~\ref{s:simulations}, this strategy is effective for intermediate
times but does not correctly promote the long periods spent near the
stable fixed point in the true dynamics.  This manifests in reduced
efficiency of the IS strategy in these cases.  To address this
phenomenon, we turn off the biasing for values of $x$ within a
diffusion length of the stable fixed point, which leads to a
significant improvement of sampling efficiency.

\section{Micromagnetic framework and thermally induced switching} \label{s:micro}

We consider a region $\Omega\subset \mathbb{R}^3$ occupied by a
ferromagnetic film with in-plane shape $D\subset \mathbb{R}^2$ and
thickness $d$, i.e., $\Omega = D \times (0,d)$, characterized by
saturation magnetization $M_s$, exchange stiffness $A$ and an in-plane
uniaxial anisotropy constant $K_u$, at temperature $k_B T$ in energy
units. To model the free layer of an in-plane STT-MRAM cell, we use
the stochastic Landau-Lifshitz-Gilbert equation
\cite{garcia07,brataas12} for the unit magnetization vector
$\mathbf{m} = (m_x,m_y,m_z)$, written in the Landau-Lifshitz form:
\begin{align}
  \frac{\partial{\bf{m}}}{\partial t} = -{\bf{m}}\times{\bf{h}} -
  \alpha{\bf{m}}\times {\bf{m}}\times{\bf{h}} +
  \alpha{\bf{m}}\times\boldsymbol{\tau}_{\mathrm{STT}}
  +\boldsymbol{\tau}_{\mathrm{STT}},
\label{eq:LL}
\end{align}
where time and lengths are measured in the units of
$\tau_0 = (1+\alpha^2) /(\gamma \mu_0 M_s)$ and
$l_{ex} = \sqrt{2 A / (\mu_0 M_s^2)}$, respectively, $\alpha$ is
Gilbert damping, $\gamma$ is the gyromagnetic ratio and
$\boldsymbol{\tau}_{STT}$ is the spin-transfer torque. The effective
field is given by
\begin{align}
  \label{heta}
{\bf{h}} = -\frac{\delta E}{\delta {\bf{m}}} +
  \sqrt{\sigma}\boldsymbol{\eta},
\end{align}
where
\begin{align}
  E[\mathbf m] = \frac 12 \int_{D}\left(
  |\nabla{\bf{m}}|^2 + Q m_y^2 + m_z^2
  \right) \,d^2r
\label{E:nonuniform}
\end{align}
is the leading order thin film micromagnetic energy measured in the
units of $2 A d$, and $Q = 2 K_u / (\mu_0 M_s^2)$ is the quality
factor \cite{gioia97}. The thermal fluctuation term
$\boldsymbol{\eta}({\bf{r}},t)$ is a delta-correlated, suitably
regularized three-dimensional spatiotemporal Gaussian white noise
\cite{hairer09}, with noise strength
\begin{align}
  \sigma =  \frac{\alpha k_BT}{Ad(1+\alpha^2)}.
\end{align}
by the fluctuation-dissipation theorem
\cite{brown63,garcia-palacios98}. Equation (\ref{eq:LL}) is to be
interpreted in the Stratonovich sense, so that it preserves the norm
constraint $|\mathbf{m}| = 1$ \cite{garcia-palacios98,garcia07}.

The spin-transfer torque $\boldsymbol{\tau}_{\mathrm{STT}}$ is given
by
\begin{align} \label{term:stt} \boldsymbol{\tau}_{\mathrm{STT}} =
  a_J{\bf{m}}\times {\bf{m}} \times{\bf{m}}_p + b_J{\bf{m}}
  \times{\bf{m}}_p .
\end{align}
where $a_J =-{\eta j \hbar}/({2d e \mu_0 M_s^2})$ and
$b_J = \beta a_J$ are dimensionless Slonczewski and field-like torque
strengths \cite{brataas12}. Here $j$ is the density of electric
current passing perpendicularly through the film, $e$ is the
elementary charge, $\eta \in(0,1]$ is the spin polarization
efficiency, $\beta$ is the relative strength of the field-like spin
torque, and ${\bf{m}}_p$ is the spin-polarization direction. In this
paper, we consider ${\bf{m}}_p = (1,0,0)^T$, i.e., when the spin
current is polarized along the easy axis in the film plane, as is the
case in the basic in-plane spin valve
\cite{Khvalkovskiy2013,pinna13jap,Newhall_Eric}.

The macrospin approximation assumes spatial uniformity across the
ferromagnet, such that the first term in (\ref{E:nonuniform}) is zero
and (\ref{eq:LL}) is an ordinary differential equation. In this case
$E(\mathbf m) = \frac12 (Q m_y^2 + m_z^2) S$, where $S$ is the area of
$D$ in the units of $l_{ex}^2$, and the effective field is given by
\begin{align}
  \label{eq:hmacro}
  {\bf{h}} = -S^{-1}\nabla_{\mathbf m} E +
  \sqrt{\frac{2\alpha \epsilon}{1+\alpha^2}} \,
  \boldsymbol{\dot{W}}(t),
\end{align}
where $\epsilon = k_B T / (2AdS)$, and $\boldsymbol{W}(t)$ 
is a three-dimensional Brownian motion. The noise
coefficient $\sqrt{{2\alpha \epsilon}/(1+\alpha^2)}$ is consistent
with the Gibbs distribution, in which $\epsilon$ plays the role of the
dimensionless temperature.

When the dimensionless parameters satisfy
\begin{align}
  \label{eq:Qsmall}
  \alpha \sim 1 \ \text{and}  \  a_J \sim b_J \sim Q \ll 1,
\end{align}
i.e., in soft materials with relatively high damping and low spin
torques, the magnetization is always constrained to lie almost
entirely in the film plane \cite{kohn05,garcia01,Lund2016}. In this
case the system (\ref{eq:LL}) may be simplified to an equation for the
angle $\theta$ such that
$\mathbf m \simeq (\cos \theta, \sin \theta, 0)$ (see Appendix):
\begin{align}  \label{sys:macro}
\dot{\theta} =  b(\theta) +{\frac{1}{\sqrt{\Delta}}}\dot{W},
\end{align}
where
\begin{equation}
b(\theta) = (I_J-\cos(\theta))\sin(\theta),
\label{e:MSdrift}
\end{equation}
$ I_J = b_J/ Q$, $ 1/\Delta = 2 \epsilon / Q$, and the unit of time is
now $\tau_Q = \alpha/(\gamma \mu_0 M_s Q)$.  We point out that even
though \eqref{sys:macro} was obtained for an in-plane STT-MRAM cell,
exactly the same equation arises in the modeling of perpendicular
cells \cite{Butler2012}. Therefore, a direct comparison with the
results obtained by Butler et al. \cite{Butler2012}, who used the
analysis of the Fokker-Planck equation is also possible.

Equation \eqref{sys:macro} with $0 \leq I_J < 1$ will be used as the
simplest example of a stochastic micromagnetic model with bistability,
for which several IS biasing strategies will be illustrated. A more
realistic model of an STT-MRAM cell would need to incorporate spatial
heterogeneity within the cell, which may be captured by considering a
system of $N$ exchange-coupled macrospins associated with the
magnetization in each of the polycrystalline grains. If each grain has
a dimensionless area $S_i$, $1/\Delta_i = k_B T / (Ad S_i Q)$, and
$\theta_i$ are such that the magnetization in each grain is
$\mathbf m_i \simeq (\cos \theta_i, \sin \theta_i, 0)$, then
(\ref{sys:macro}) may be generalized to
\cite{miles91,miles95,fidler00}
  \begin{align}
    \label{thetai}
    \dot{\theta}_i =  b(\theta_i) + \sum_{j = 1}^N a_{ij} S_i^{-1}
    \sin(\theta_j - \theta_i)
    +{\frac{1}{\sqrt{\Delta_i}}} \, \dot{W}_i,
  \end{align}
  using Heisenberg exchange with dimensionless strengths
  $a_{ij} = a_{ji} \geq 0$ for the interactions between the grains,
  with $W_i$ being $N$ uncorrelated Brownian motions. Note that this
  equation is a stochastic version of a gradient system governed by an
  {\em effective potential}
  \begin{multline}
    \label{VN}
    V_N(\theta_1, \ldots, \theta_N) = \sum_{i=1}^N S_i \left( I_J \cos
      \theta_i + \tfrac12
      \sin^2 \theta_i \right) \\
    - \sum_{i=1}^{N-1} \sum_{j=i+1}^N a_{ij} \cos(\theta_i -
    \theta_j),
  \end{multline}
  and obeying detailed balance. The coupling coefficients $a_{ij}$ are
  non-zero only for the nearest neighbors and may in principle be
  determined from the geometric characteristics of the individual
  grains. For simplicity, in this paper we will limit ourselves to the
  consideration of the case of two identical macrospins only, which
  may correspond, e.g., to exchange-coupled synthetic bilayers
  \cite{taniguchi11}.

  The single macrospin drift term defined in (\ref{e:MSdrift}) has
  stable fixed points $\theta = 0$ and $\theta = \pm \pi$, separated
  by unstable fixed points $\theta = \pm \theta_J$, where
  $ \theta_J =\arccos(I_J) \leq \pi / 2$. Therefore, for a trajectory
  starting close to $\theta = 0$ at $t = 0$ a switching event for the
  time horizon $T > 0$ would be defined as one in which $|\theta|$ is
  close to $\pi$ at $t = T$. For the purposes of this paper, we
  consider a switching event to have occurred if
  $|\theta(T_{sw})| = \pi/2$ for some $0 < T_{sw} \leq T$, starting
  with $\theta(0) = 0$, i.e., the macrospin changes its direction
  along the easy axis. Similarly, for the system of two coupled
  macrospins governed by \eqref{thetai} and starting with
  $\theta_1(0) = \theta_2(0) = 0$, we define a switching event to have
  occurred if $\max(|\theta_1(T_{sw})|, |\theta_2(T_{sw})|) = \pi/2$
  for some $0 < T_{sw} \leq T$, i.e., at least one macrospin changes
  its direction along the easy axis.


  The exact finite time switching probability for a single macrospin
  exhibited by~(\ref{sys:macro}) can be computed by solving the
  backward Fokker-Planck equation for the probability
  $P_{\mathrm{sw}}(\theta, t)$ that a trajectory starting at a given value of
  $\theta \in (-\pi/2, \pi/2)$ reached the value of
  $\theta = \pm \pi/2$ by time $t$ \cite{gardiner2004handbook}. This
  probability satisfies
\begin{equation} \label{bfp:1} \frac{\partial P_{\mathrm{sw}}}{\partial t} =
  b(\theta)\frac{\partial P_{\mathrm{sw}} }{\partial \theta} + \frac{1}{2
    \Delta} \frac{\partial^2 P_{\mathrm{sw}}}{\partial \theta^2},
\end{equation}
for $(\theta, t) \in (-\pi/2, \pi/2) \times (0, T)$, with initial and
boundary conditions
\begin{equation}
  P_{\mathrm{sw}}(\theta,0) = 0 , \qquad P_{\mathrm{sw}}(\pm \pi/2, t) = 1,
\end{equation}
respectively.  In particular, the probability of having switched by
time $T$, given the initial state $\theta(0)=0$ is equal to
$P_{\mathrm{sw}}(0,T)$.

As an alternative to solving \eqref{bfp:1}, we incorporate IS in the
estimation of $P_{\mathrm{sw}}$ by sampling controlled dynamics
\begin{equation}
  \label{eq:thmonoW}
  \dot{\tilde\theta}  = b(\tilde\theta) + u^* +
  \frac{1}{\sqrt{\Delta}} \dot{W}, \qquad \tilde\theta(0) = 0,
\end{equation}
for $t \in (0, T)$.  For finite-time bias we have $u^* = u^*_T$, where
$u_T^*(\tilde \theta,t) = \dot \theta^T_{\tilde\theta, t}(t) -
b(\tilde \theta)$ is the bias function obtained through
\eqref{control:1} by interpreting \eqref{eq:thmonoW} as an
  instance of \eqref{eqn:controlsystem} with
  $\epsilon = 1/\sqrt{\Delta}$, using the minimizer
  $\theta^T_{\tilde \theta, t}(s)$ of the finite-time action
\begin{equation}
  S_T[\theta] = \int_t^T \frac{1}{2}|
  \dot{\theta}(s) - b(\theta(s))|^2 ds,
\end{equation}
among all $\theta(s)$ with $\theta(t) = \tilde \theta$ and
$|\theta(T)| = \pi/2$. For sufficiently large time horizons, we set
instead $u^* = u^*_\infty$, where the ininfite-time bias $u^*_\infty$
is obtained from the minimizer of the infinite-time action
$S_{\infty}$. In the single macrospin case the latter is simply given
by a straight line segment, resulting in a particularly simple
explicit form of the bias:
\begin{equation}
  \label{control:u}
  u_\infty^*(\tilde\theta)
  = \begin{cases}
    -2b(\tilde\theta), & -\theta_J \leqslant \tilde\theta
    \leqslant \theta_J, \\ 
    0, & \text{otherwise}.
  \end{cases}
\end{equation}

To account for switching events, we stop the
trajectory at time $t = T_{sw}$ as soon as the switching criterion
$|\tilde\theta(t)| = \pi/2$ is satisfied, or otherwise set
$T_{sw} = T$. The likelihood ratio is recovered from
\eqref{eq:L}, which in this case is
 \begin{align}
   \label{eq:Lmacro}
   L = \exp \Biggl(
   & -{\frac{\Delta}2} \int_0^{T_{sw}}
     |u^*(t)|^2 dt \notag \\
   & - \sqrt{\Delta} \int_0^{T_{sw}}
     u^*(t) \, dW(t) \Biggr),
 \end{align}
 where either $u^*(t) = u^*_T(\tilde \theta(t),t)$ or
 $u^*(t) = u^*_{\infty}(\tilde \theta(t))$, depending on whether we
 use the finite- or infinite-time bias function, respectively, and
 $\tilde\theta(t)$ is the solution of \eqref{eq:thmonoW} with a
 particular realization $W(t)$ of the noise.

\section{Simulations}
\label{s:simulations}

The following sections describe results obtained from IS simulations
of macrospin and coupled-spin systems using physical parameters drawn
from Ref.~\cite{Butler2012} for the purpose of comparison. All
simulations use the Euler-Maruyama method with a fixed time step
$\tau = 0.1$.

The IS results presented below are obtained using bias functions based
on either finite- or infinite-time minimizers of the action given
by~(\ref{min_Action}).  Infinite-time bias functions are based on
\eqref{control:u} in the macrospin case, and therefore do not require
additional computation.  Infinite-time bias functions for the
coupled-spin system are obtained by minimizing the action in
\eqref{min_Action} through the geometric minimum action method (GMAM)
with 50 gridpoints~\cite{heymann_geometric_2008}.  Finite-time bias
functions are obtained by minimizing the action in \eqref{min_Action}
through a combination of Newton's method for the associated
Euler-Lagrange equation and the improved adaptive minimum action
method~\cite{YSun2018}, with $500$ gridpoints in the macrospin case
and $100$ in the coupled-spin case.

\begin{figure*}[ht]
   \centering
         \subfigure{\includegraphics[width =3.52in]{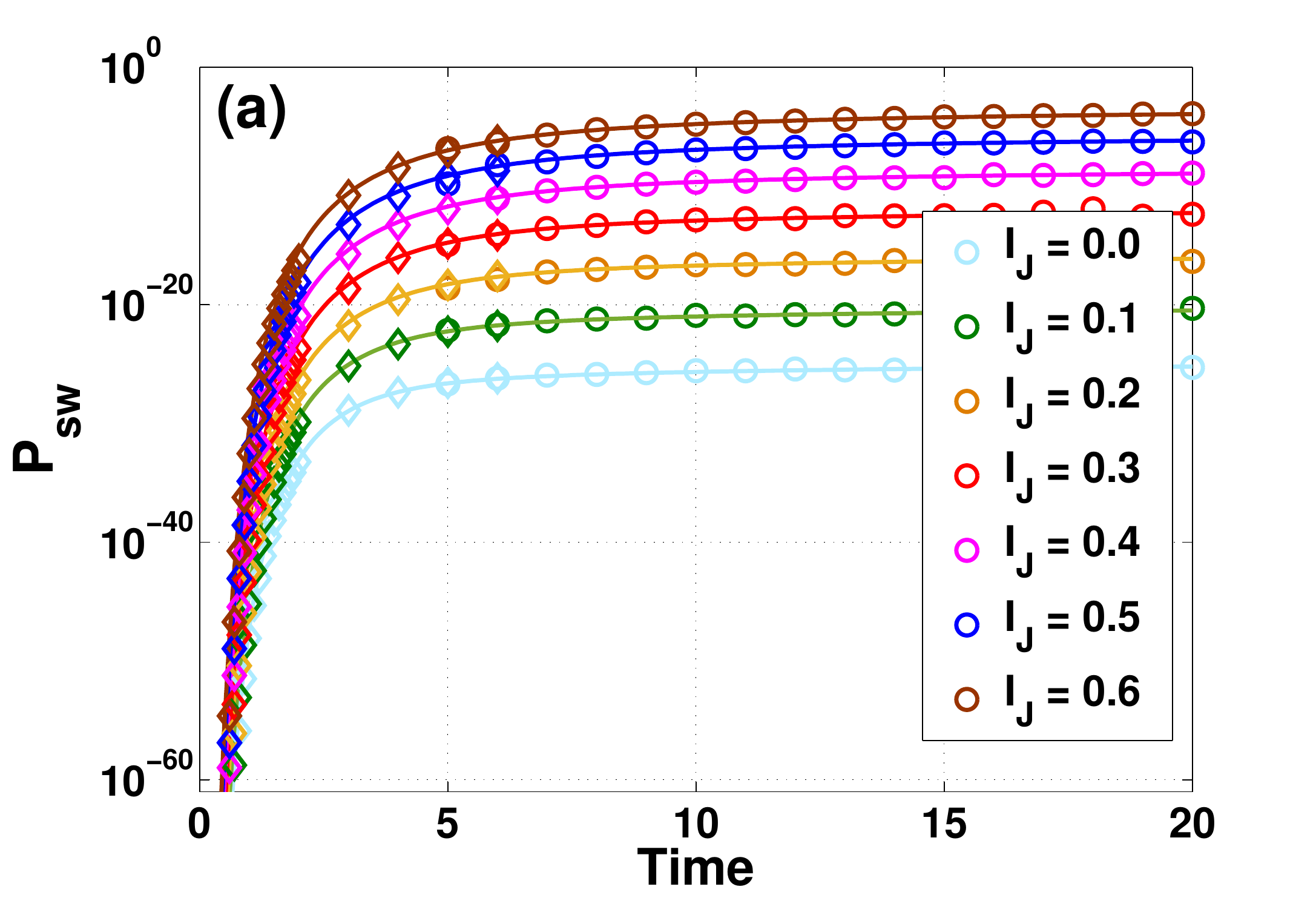}}
          \subfigure{\includegraphics[width =3.5in]{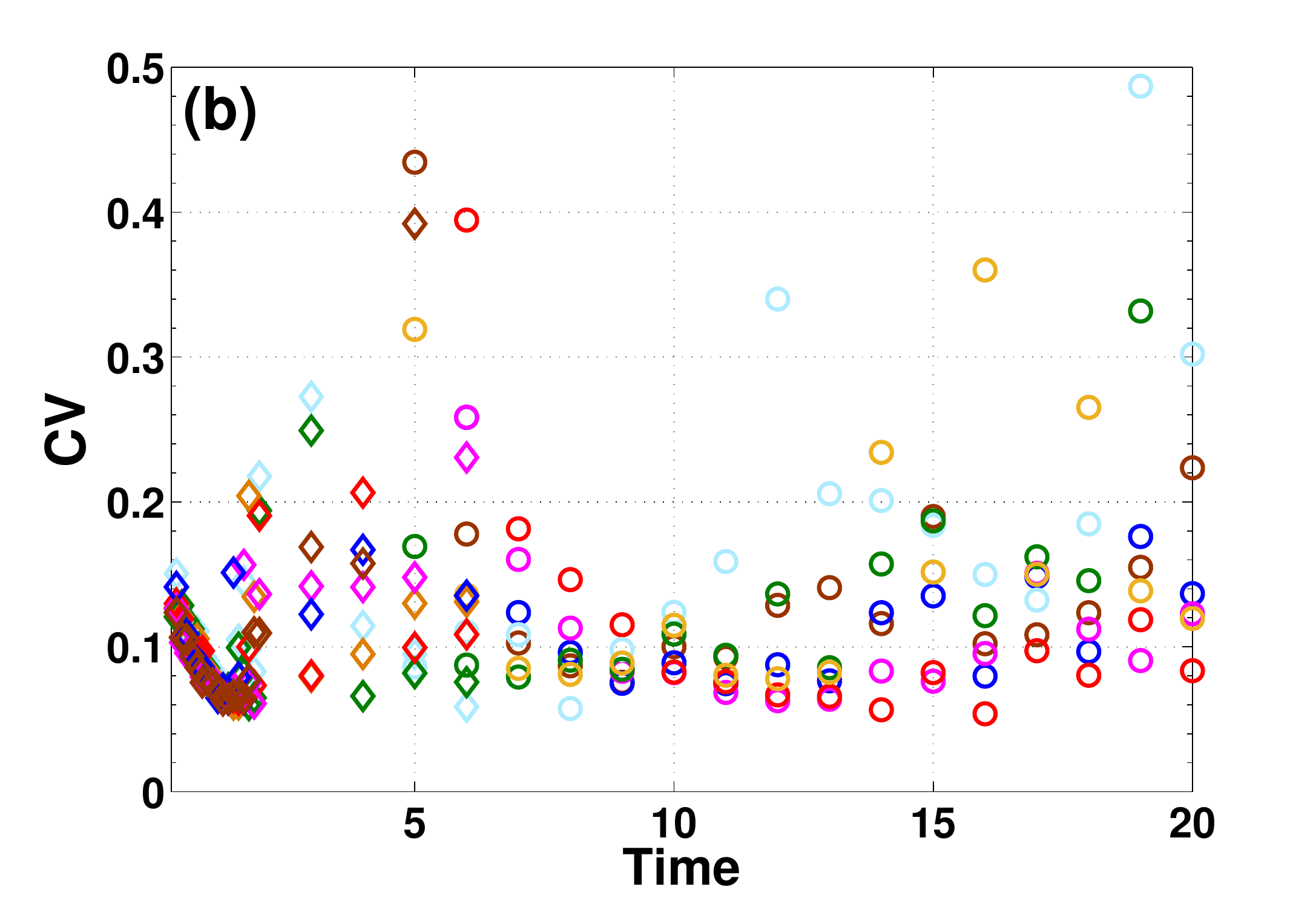}}
          \caption{IS estimate for switching probability (RSER) and CV
            of the RSER with sample size $M = 10^3$.  (a) RSER vs.\
            reading pulse duration $T$ and reading current $I_J$ with
            thermal stability factor $\Delta = 60.$ Solid lines denote
            numerical solutions of FPE~(\ref{bfp:1}).  Open circles
            and open diamonds denote estimates generated by IS with
            finite-time bias functions and infinite-time bias
            functions, respectively, color coded by current
            amplitude. For $T \ge 5$ infinite-time bias functions are
            used, while for IS and for $T \le 6$ finite-time bias
            functions are used. IS results at $T=5$ and $T=6$ obtained
            using infinite- and finite-time bias functions are
            indistinguishable. (b) CV of the RSER.}
   \label{fig:MacroIS60}
\end{figure*}

\begin{figure*}[ht]
   \centering
        \subfigure{ \includegraphics[width = 3.52in]{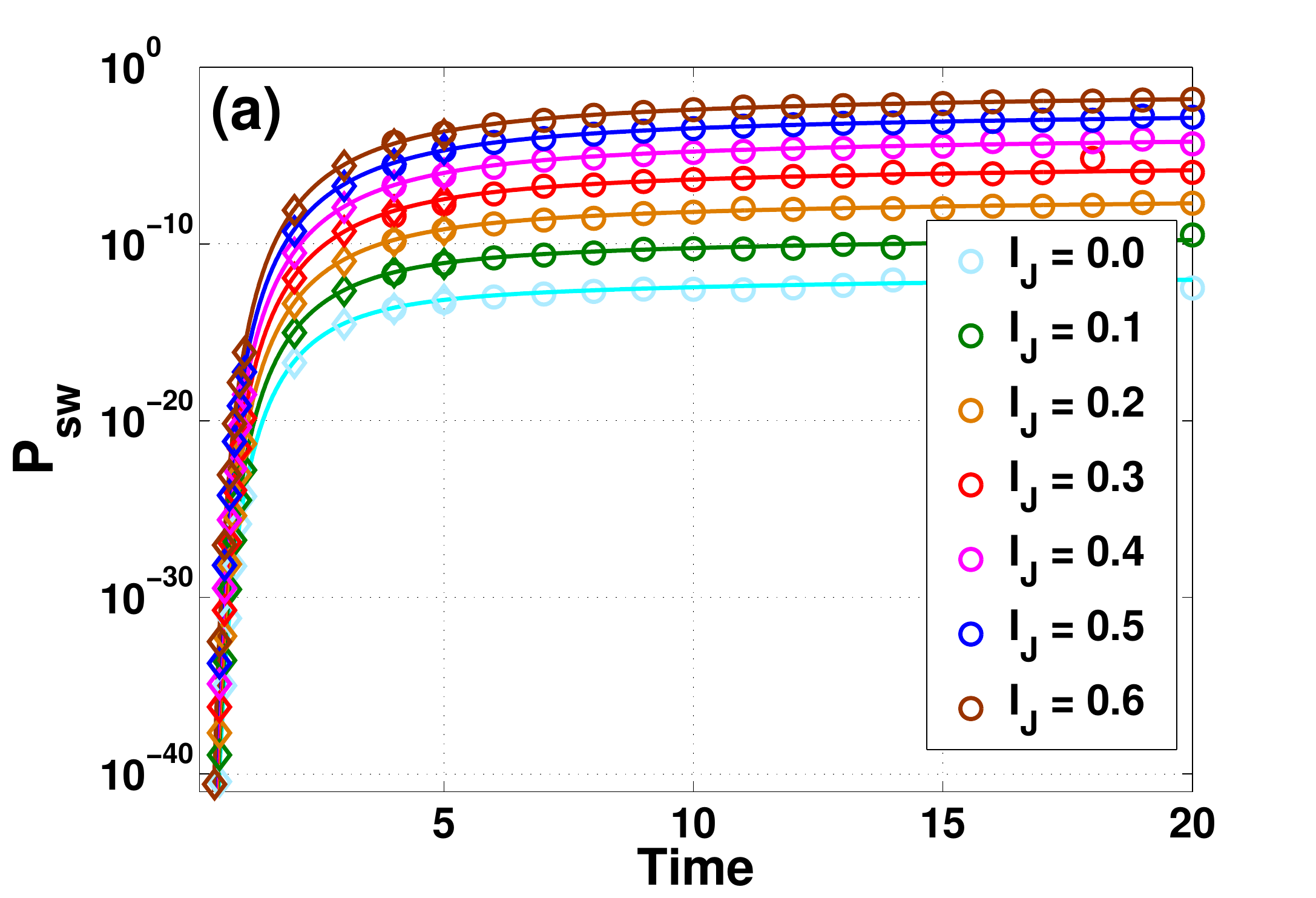}}
         \subfigure{\includegraphics[width = 3.5in]{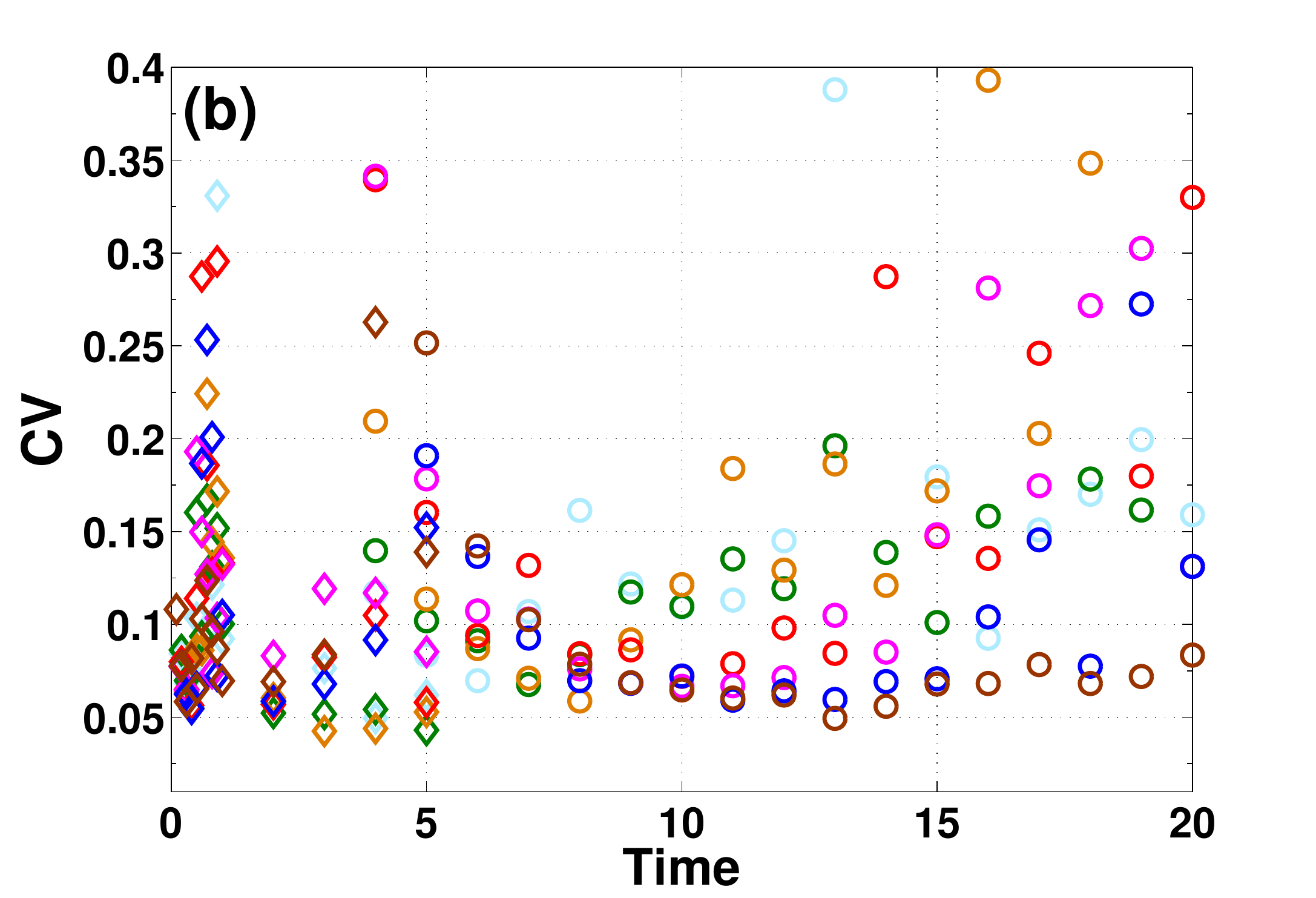}}
         \caption{Similar to Fig.~\ref{fig:MacroIS60}, but with
           $\Delta=30$. For {$T \ge 4$} infinite-time bias functions
           are used, and for {$T \le 5$} finite-time bias functions
           are used. IS estimates at $T=4$ and $T = 5$ obtained using
           infinite- and finite-time bias functions are
           indistinguishable.}
   \label{fig:MacroIS30}
\end{figure*}

\subsection{Single macrospin}

The discretized version of (\ref{eq:thmonoW}) reads explicitly
\begin{equation}
  \tilde\theta^{k+1}  =\tilde\theta^k + (b(\tilde\theta^k)
  +u^*(\tilde\theta^k, t_k)) \tau +  \frac{\sqrt{\tau}}{
    \sqrt{\Delta}}  \xi_k, \quad \tilde \theta^0 = 0,
     \label{e:thmonoWdisc}
\end{equation}
where $\tilde\theta^k = \tilde \theta(t_k)$, $t_k = k \tau$ for
$k = 0, 1, \ldots, K$, and $\xi_k$ are independent and drawn from the
standard normal distribution. The value of $K$ is chosen so that
either $|\tilde \theta_k| < \pi/2$ for all
$k < K < \lfloor T/\tau \rfloor$ and $|\tilde\theta^K| \ge \pi/2$, or
$K = \lfloor T/\tau \rfloor$, i.e., we stop the simulation if a
switching event occurs at time $t_k < T$. The likelihood ratio
corresponding to \eqref{eq:Lmacro} is then given explicitly by
\begin{multline}
  \label{Ldisc}
  L = \exp \Biggl( -{\tau \Delta \over 2} \sum_{k=0}^K
  |u^*(\tilde\theta^k, t_k)|^2   \\
  - \sqrt{\tau \Delta} \sum_{k=0}^K u^*(\tilde\theta^k, t_k)\,
  \xi_k\Biggr).
\end{multline}
The figures below are generated using switching probabilities and
their coefficients of variation computed by applying
formulas~(\ref{e:ISestimator}) and~(\ref{e:ISCV}), respectively, to
ensembles of $M$ runs, where $I(\omega_i)=1$ for runs that generate a
switching event and $I(\omega_i)=0$ otherwise.

Figures~\ref{fig:MacroIS60} and~\ref{fig:MacroIS30} show the RSER as a
function of time for seven values of $I_J$ between $0$ and $ 0.6$ with
thermal stability factors $\Delta = 60$ and $\Delta=30$, respectively.
Runs at both temperatures are included here to facilitate comparison
with the results presented in Ref.~\cite{Butler2012}.  Both show a
comparison between numerical solutions of backward FPE~(\ref{bfp:1})
and IS simulations for the macrospin model. For $T \ge 5$ in
Fig.~\ref{fig:MacroIS60} and $T\ge 4$ in Fig.~\ref{fig:MacroIS30},
infinite-time bias functions are used for IS while for $T \le 6$ in
Fig.~\ref{fig:MacroIS60} and $T\leq 5$ in Fig.~\ref{fig:MacroIS30},
finite-time bias functions are used.  Agreement between the FPE and IS
results is excellent throughout the range of times and currents, and
the IS results are internally consistent between the finite- and
infinite-time bias functions used at $T=5$ and $T=6$ in
Fig.~\ref{fig:MacroIS60}, and used at $T=4$ in
Fig.~\ref{fig:MacroIS30}.

Fig.~\ref{fig:MacroIS60} (b) shows the CV for the IS estimates in
Fig.~\ref{fig:MacroIS60} (a), and Fig.~\ref{fig:MacroIS30} (b) shows
those for the estimates shown in Fig.~\ref{fig:MacroIS30} (a). The CV
values for the IS estimates range from approximately $0.05$ to $0.5$.
We note that with an inappropriate choice of bias the CV can be an
imperfect measure of accuracy for Monte Carlo estimates using variance
reduction \cite{glasserman_counterexamples_1997}.  However, this is
precluded by our choice of an asymptotically optimal bias function
that is based on large deviation theory
\cite{dupuis_subsolutions_2007,Eric2012}, as can also be seen from the
excellent agreement with the solutions to FPE~(\ref{bfp:1}). The low
CV values obtained for such extremely small probabilities with
moderate sample size of $M=10^3$ are therefore a clear demonstration
of the efficiency of the bias functions used here.

\begin{figure}[t]
   \centering
         \includegraphics[width = 3.6in]{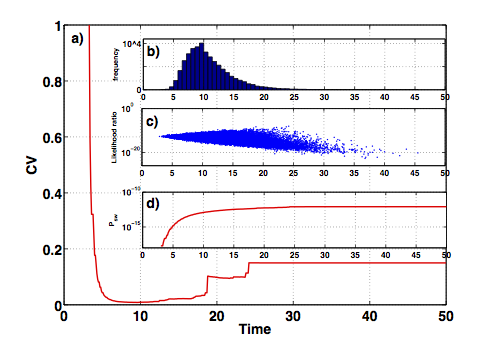}
   \caption{CVs of IS estimators for $I_J = 0.3$ with infinite-time bias functions and sample size $10^5$. Inset (a) is a histogram of exit times for the biased system. Inset (b) is the estimated likelihood ratio vs.\ time. Inset (c) is the probability of switching before time $T$. The sample size here is $M=10^5$.}
   \label{fig:Two_tialsofCV_I_J_03}
\end{figure}

In both figures the CVs are observed to increase when $T$ is decreased
from $T=10$ to $T=5$, indicating that the infinite-time bias function
used for these runs becomes progressively less efficient at capturing
the switching events as the time horizon shrinks.  The application of
finite-time bias functions for smaller times lowers the CVs as
expected.
Furthermore, Figs.~\ref{fig:MacroIS60} (b) and~\ref{fig:MacroIS30} (b)
show a similar pattern in the CVs generated by IS with infinite-time
bias functions, where the CVs also increase for {\em large\/} times
$T$. Fig.~\ref{fig:Two_tialsofCV_I_J_03} illustrates this sudden
increase in CV in the context of longer times, as well as a histogram
of switching times, and the time evolution of the spread in values of
likelihood ratio.

\begin{figure}[t]
   \centering
         \includegraphics[width = 3.6in]{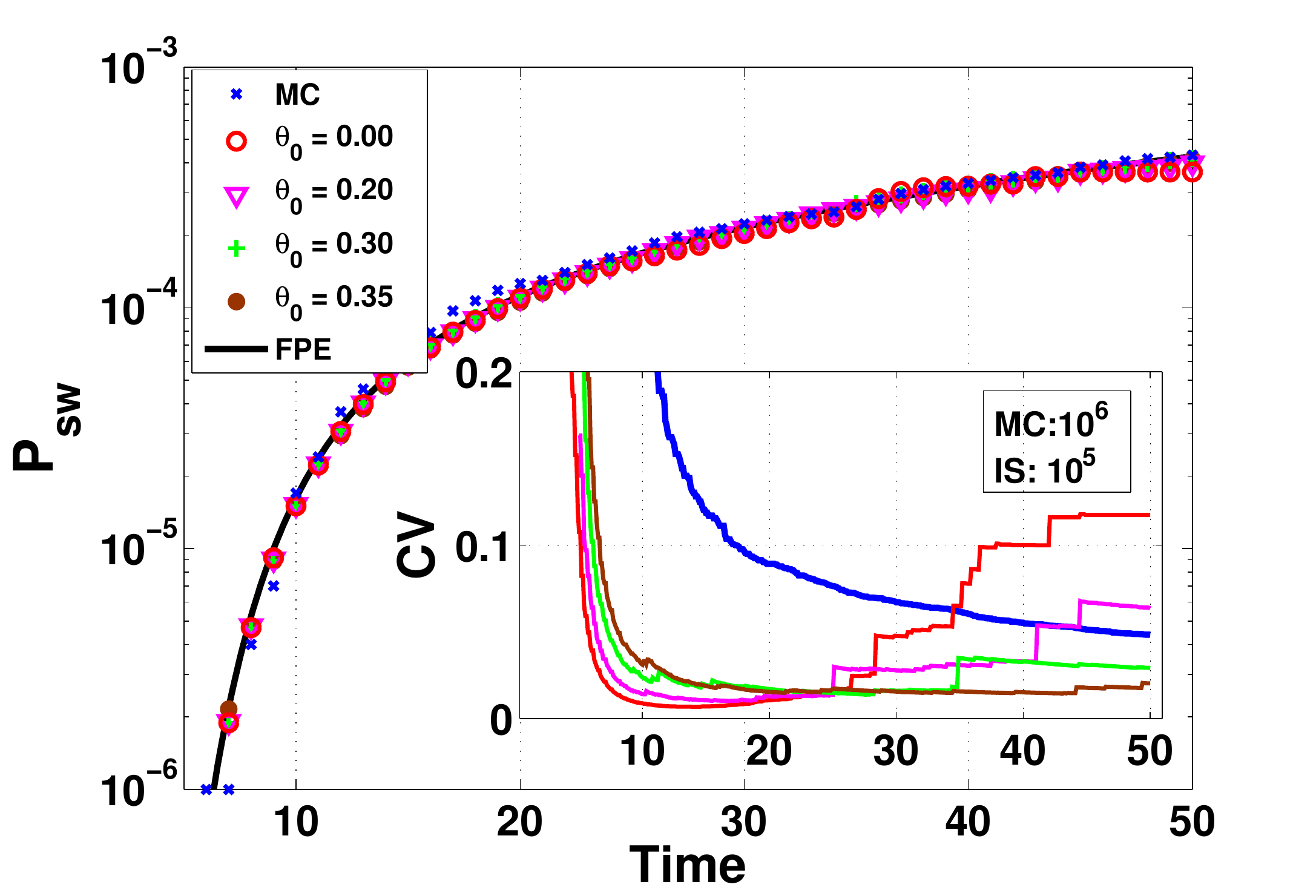}
         \caption{Switching probabilities and CVs of IS estimators for
           $I_J = 0.6$ with infinite-time bias functions active
           outside of a region containing the stable fixed point,
           defined by $\theta_0 <|\theta|<\theta_J$, for different
           values of $\theta_0$. The sample size is $10^5$.}
   \label{fig:CVvsCutoff}
 \end{figure}

This decrease in efficiency of the infinite-time bias function for
large but finite time horizons is due to the fact that exits have a
natural finite time scale dictated by diffusion near the fixed points
with the action minimizer bridging the gap between them.  When the
time horizon of the simulation is large relative to this time scale,
it allows for exit events that hover near the stable fixed point
before exiting just prior to the horizon time.  These events occur
with considerably higher likelihood under the {\em unbiased\/}
dynamics than under the biased dynamics, leading to a very large
likelihood ratio that causes them to dominate the CV computation.
Since the diffusion time grows as the noise strength decreases, this
phenomenon can be regarded as a finite-noise effect, and it indeed
vanishes as $\Delta\to\infty$.  We address this issue for finite noise
by turning off the biasing near the stable fixed point, i.e., for
$|\theta|< \theta_0$, with the results plotted for different values of
$\theta_0$ in Fig.~\ref{fig:CVvsCutoff}.  It is clearly seen that as
$\theta_0$ increases, the anomalous behavior for large horizon time is
mitigated, at the expense of sampling efficiency for small horizon
times.
%

\subsection{Two coupled identical macrospins: model}

To demonstrate that the IS method can also be effective in coupled
systems, we simulate two spins with identical volume and dynamics
given by \eqref{thetai}, which in this case is explicitly
\begin{align}
  \dot{\theta}_1 &=c \sin( \theta_{2} - \theta_{1} )
                   +b(\theta_1) +
                   \sqrt{\frac{2}{{\Delta}}}\dot{W_1}, \label{th1} \\
  \dot{\theta}_2 &=c \sin( \theta_{1} - \theta_{2} )
                   +b(\theta_2)+ \sqrt{
                   \frac{2}{{\Delta}}}\dot{W_2}, \label{th2}
 \end{align}
 where $c > 0$ is the ferromagnetic exchange coupling strength
 favoring parallel alignment of the two spins. The initial conditions
 are $\theta_1(0) = \theta_2(0) = 0$. Recall that as a switching
 criterion we adopt that at least one of the angles reaches $\pi/2$ in
 absolute value.

 At finite temperature, the rare events of switching for the coupled
 spin system occur along the maximum likelihood paths, which are also
 the minimum energy paths of the system. When the system undergoes a
 transition, it switches by coherent rotation for strongly coupled
 spins, asymmetric coherent rotation for weakly coupled spins, or
 single particle reversal for extremely weakly coupled
 spin~\cite{chen1992}. More precisely, in the limit of infinite
 coupling strength $c$, the coupled spin system collapses to the
 macrospin model with $\theta_1=\theta_2$, and the most probable path
 terminating at $\max(|\theta_1|,|\theta_2|)=\pi/2$ is identical to
 that of a single macrospin.  As the coupling strength decreases, the
 dynamics of the coupled spin system changes significantly, ultimately
 leading to spins that evolve independently. Note that in
 $N$-dimensional coupled spin systems, the sequence of bifurcations
 from single macrospin dynamics to $N$-fold macrospin dynamics as $c$
 decreases from infinity~\cite{berglund_metastability_2007} further
 exacerbates the challenge in finding appropriate bias functions.

 \subsection{Two coupled identical macrospins: biased dynamics}

 The biased dynamics associated with \eqref{th1} and \eqref{th2} reads
 \begin{align}
   \dot{\tilde \theta}_1 &=c \sin( \tilde \theta_{2} - \tilde \theta_{1} )
                           +b(\tilde \theta_1) + \sqrt{2} \, u^*_1    + 
                           \sqrt{\frac{2}{{\Delta}}} \dot{W_1},\\
   \dot{\tilde \theta}_2 &=c \sin( \tilde \theta_{1} - \tilde \theta_{2} )
                           +b(\tilde \theta_2)+ \sqrt{2} \, u^*_2 +
                           \sqrt{\frac{2}{{\Delta}}} \dot{W_2}.
 \end{align}
 In contrast to the single macrospin case, for two coupled macrospins
 an exact analytical bias function is no longer available even for
 infinite-time biasing. Therefore, it is necessary to obtain numerical
 bias functions by minimizing the Freidlin-Wentzell
 action~(\ref{min_Action}) with terminal condition
 $\max(|\theta_1(T_{sw})|, |\theta_2(T_{sw})|) = \pi/2$ for some
 $0 < T_{sw} \leq T \leq \infty$.

 For finite-time bias the action functional $S_T$ is given explicitly
 by
 \begin{multline}
   \label{ST2}
   S_T[\theta_1, \theta_2] = \frac14 \int_t^T \left( \dot \theta_1 -
     c \sin (\theta_2 - \theta_1) - b(\theta_1) \right)^2 ds \\
   + \frac14 \int_t^T \left( \dot \theta_2 - c \sin (\theta_1 -
     \theta_2) - b(\theta_2) \right)^2 ds,
 \end{multline}
 and the corresponding finite-time bias
 $(u^*_1, u^*_2) = (u^*_{T,1}, u^*_{T,2})$ is
 \begin{align}
   u^*_{T,1}(\tilde \theta_1, \tilde \theta_2, t)  =
   \frac{1}{\sqrt{2}} ( \dot \theta^T_{\tilde \theta_1, \tilde
   \theta_2, t, 1}(t) - c \sin (\tilde \theta_1 - \tilde \theta_2) -
   b(\tilde \theta_1)), \\
   u^*_{T,2}(\tilde \theta_1, \tilde \theta_2, t)  =
   \frac{1}{\sqrt{2}} ( \dot \theta^T_{\tilde \theta_1, \tilde
   \theta_2, t, 2}(t) - c \sin (\tilde \theta_2 - \tilde \theta_1) -
   b(\tilde \theta_2)),
 \end{align}
 where
 $\theta^T_{\tilde \theta_1, \tilde \theta_2, t}(s) = (\theta^T_{
   \tilde \theta_1, \tilde \theta_2, t, 1}(s), \theta^T_{\tilde
   \theta_1, \tilde \theta_2, t, 2}(s))$ is the minimizer of $S_T$
 satisfying
 $\theta^T_{\tilde \theta_1, \tilde \theta_2, t}(t) = (\tilde
 \theta_1, \tilde \theta_2)$ and
 $\max ( |\theta^T_{\tilde \theta_1, \tilde \theta_2, t, 1}(T)|,
 |\theta^T_{\tilde \theta_1, \tilde \theta_2, t, 2}(T))| = \pi/2$.
   
 For infinite-time bias, we minimize
 \begin{multline}
   \label{Sinf2}
   S_\infty[\theta_1, \theta_2] = \frac12 \int_0^1 \biggl(
   \lambda(\theta_1, \theta_2) \sqrt{
     |\theta'_1|^2 + |\theta'_2|^2} \\
   - (c \sin (\theta_2 - \theta_1) + b(\theta_1)) \theta_1' - (c \sin
   (\theta_1 - \theta_2) + b(\theta_2)) \theta_2' \biggr) ds,
 \end{multline}
 where
 \begin{multline}
   \label{lam}
   \lambda(\theta_1, \theta_2) = \Big[ (c \sin (\theta_2 - \theta_1) \\
   + b(\theta_1))^2 + (c \sin (\theta_1 - \theta_2) + b(\theta_2))^2
   \Big]^{1/2},
 \end{multline}
 and express the bias as
 \begin{align}
   \label{uinf2}
   u^*_{\infty,1} & (\tilde\theta_1, \tilde\theta_2)
                    = \frac{1}{\sqrt{2}} \notag
   \\
                  & \times \left(
                    { \lambda(\tilde\theta_1, 
                    \tilde\theta_2) \theta'_{\tilde\theta_1,
                    \tilde\theta_2, 1}(0) \over 
                    \sqrt{|\theta'_{\tilde\theta_1, \tilde\theta_2,1}|^2 +
                    |\theta'_{\tilde\theta_1, \tilde\theta_2,2}|^2} } - c \sin 
                    (\tilde\theta_2 - \tilde\theta_1) -
                    b(\tilde\theta_1) \right), 
   \\
   u^*_{\infty,2} & (\tilde\theta_1, \tilde\theta_2)
                    = \frac{1}{\sqrt{2}} \notag
   \\
                  & \times \left(
                    { \lambda(\tilde\theta_1, 
                    \tilde\theta_2) \theta'_{\tilde\theta_1,
                    \tilde\theta_2, 2}(0) \over 
                    \sqrt{|\theta'_{\tilde\theta_1, \tilde\theta_2,1}|^2 +
                    |\theta'_{\tilde\theta_1, \tilde\theta_2,2}|^2} } - c \sin 
                    (\tilde\theta_1 - \tilde\theta_2) -
                    b(\tilde\theta_2) \right),  
 \end{align}
 where
 $\theta_{\tilde \theta_1, \tilde \theta_2} = (\theta_{\tilde
   \theta_1, \tilde \theta_2, 1}, \theta_{\tilde \theta_1, \tilde
   \theta_2, 2})$ is the minimizer of $S_\infty$ with
 $\theta_{\tilde \theta_1, \tilde \theta_2, t}(0) = (\tilde \theta_1,
 \tilde \theta_2)$ and
 $\max ( |\theta_{\tilde \theta_1, \tilde \theta_2, t, 1}(1)|,
 |\theta_{\tilde \theta_1, \tilde \theta_2, t, 2}(1))| = \pi/2$.
 Finally, the discretized version of the biased equations and the
 likelihood ratio are straightforward generalizations of
 \eqref{e:thmonoWdisc} and \eqref{Ldisc}.

\begin{figure*}[ht]
   \centering
          \subfigure{\includegraphics[width =3.4in]{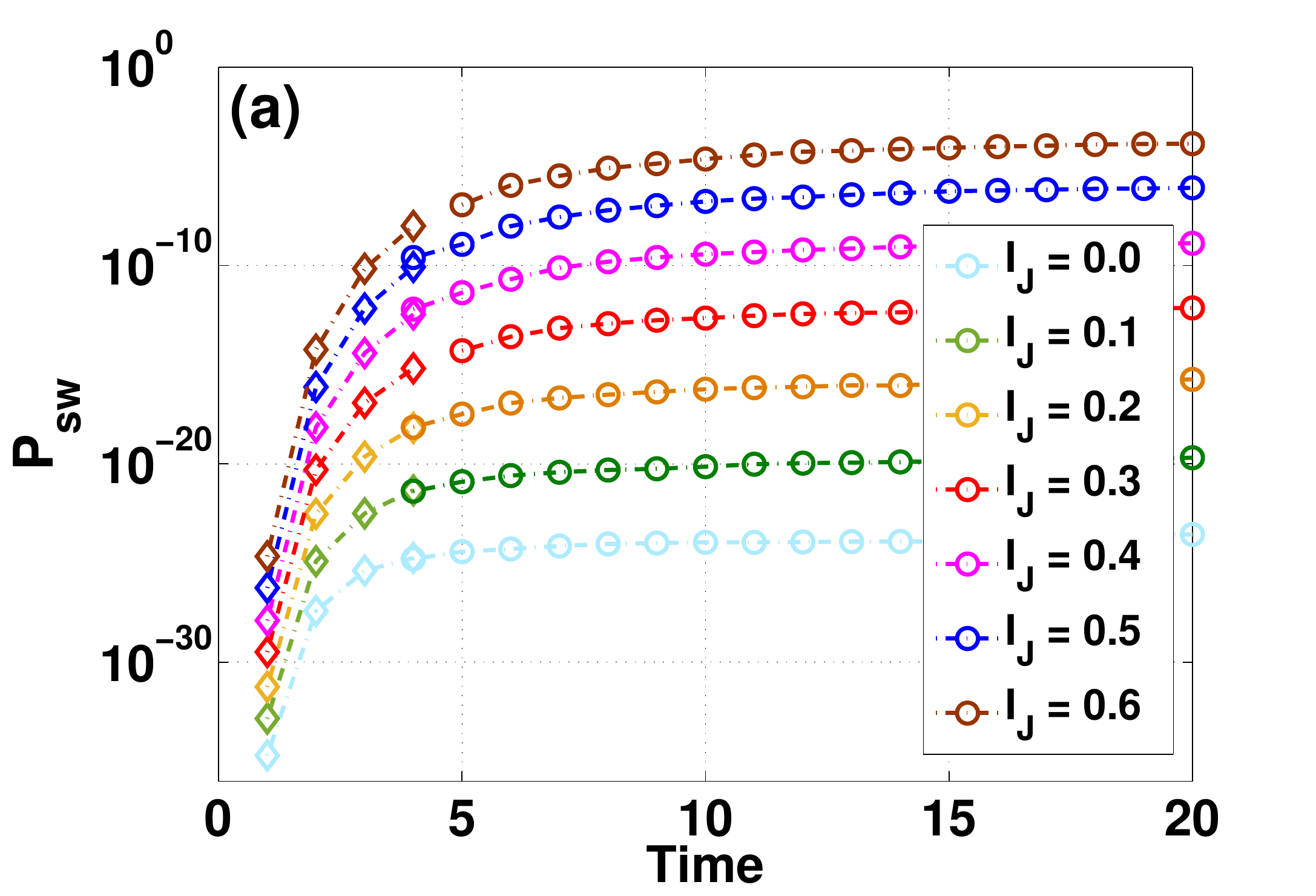}}
          \subfigure{\includegraphics[width =3.4in]{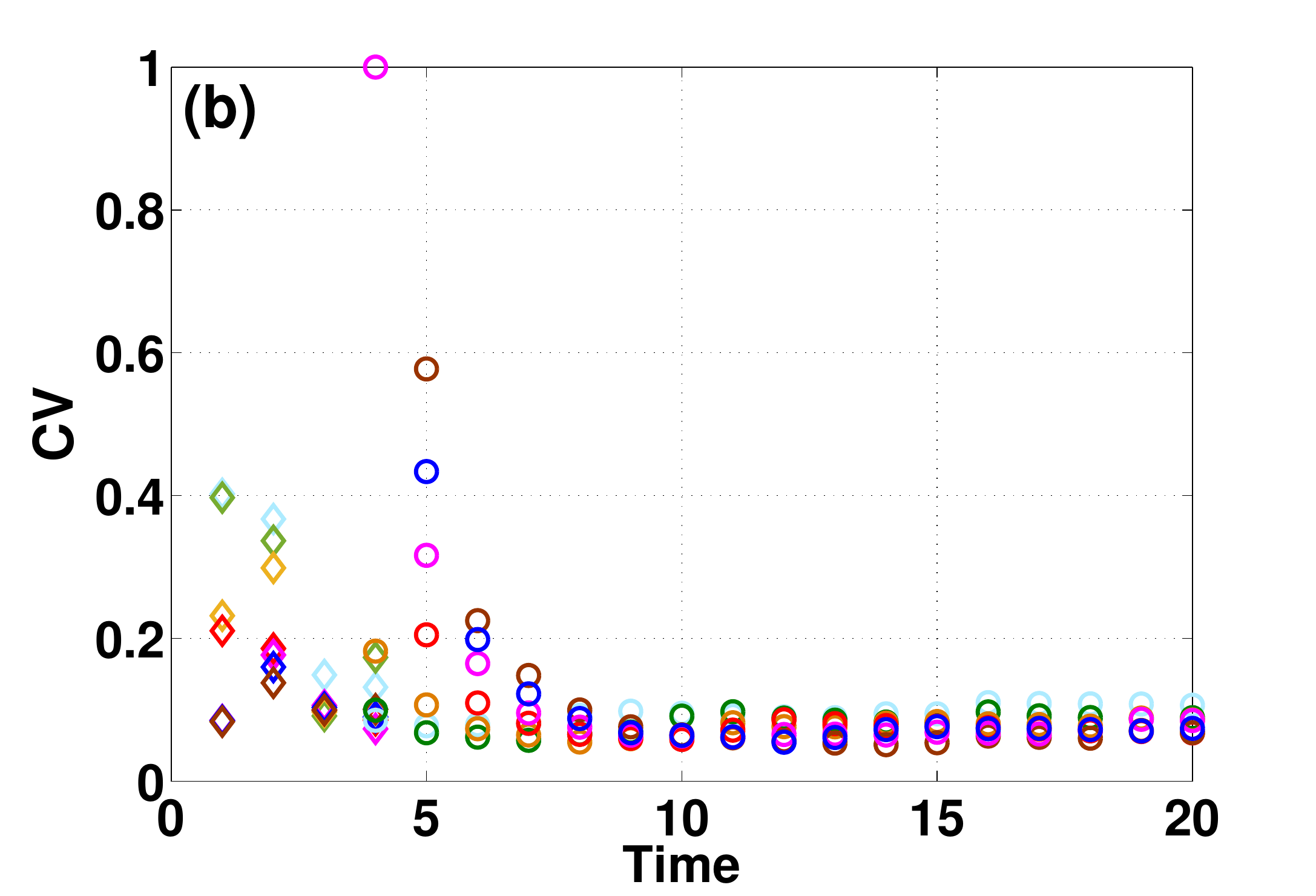}}
          \caption{IS estimate for switching probability (RSER) and CV
            of the RSER with sample size $M = 10^3$ for a strongly
            coupled two-spin system. (a) RSER with applied currents
            ranging from $0$ to $0.6$ versus time $T$ with thermal
            stability factor $\Delta = 60$ and coupling strength
            $ c = 0.8$. (b) CVs of IS estimates in (a). In both
            panels, open circles and open diamonds denote estimates
            generated by IS using infinite-time and finite-time bias
            functions, respectively. The colors correspond to
            different current amplitudes indicated in (a).}
   \label{figs:ISvsMC1_short}
\end{figure*}

\begin{figure*}[ht]
   \centering
          \subfigure{\includegraphics[width =3.4in]{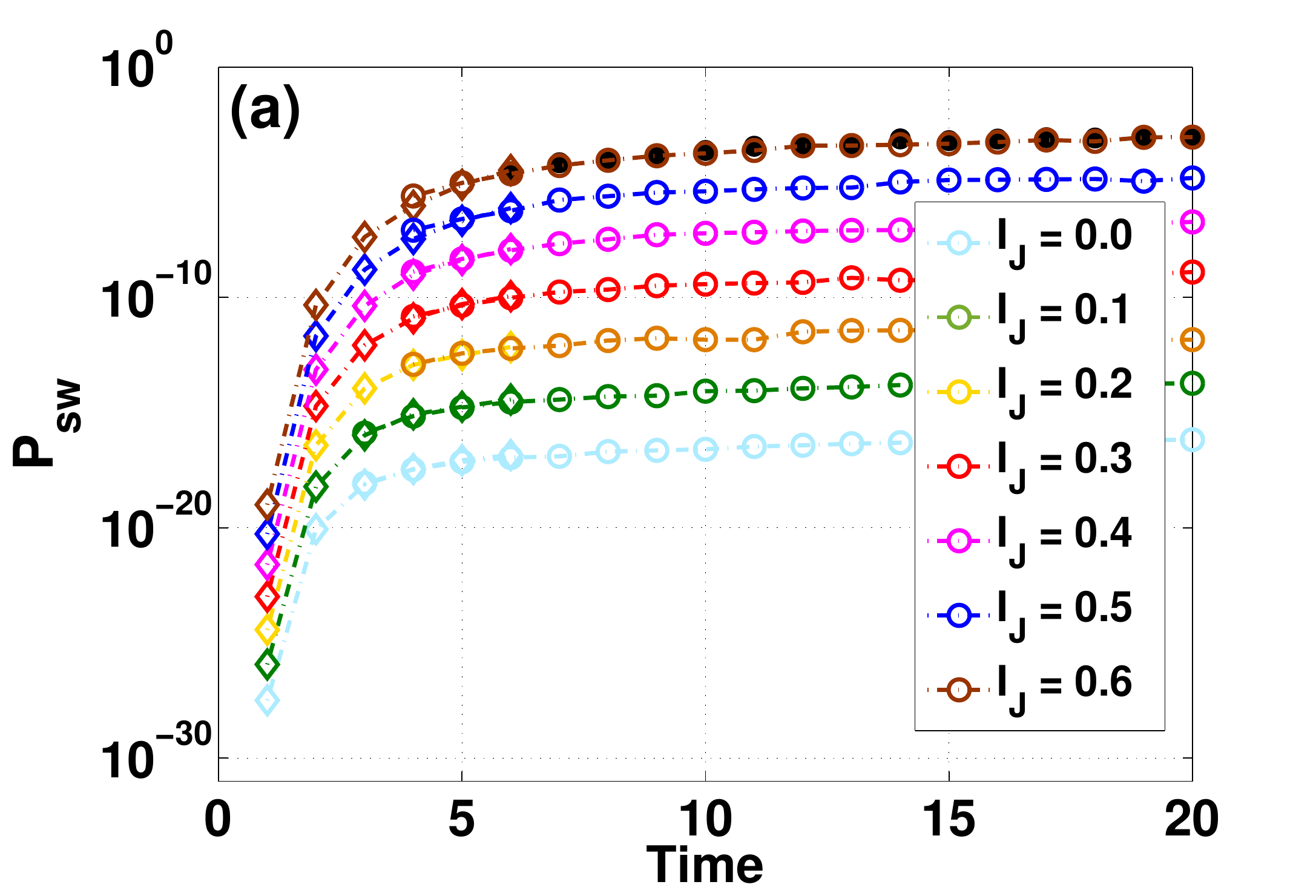}}
          \subfigure{\includegraphics[width =3.4in]{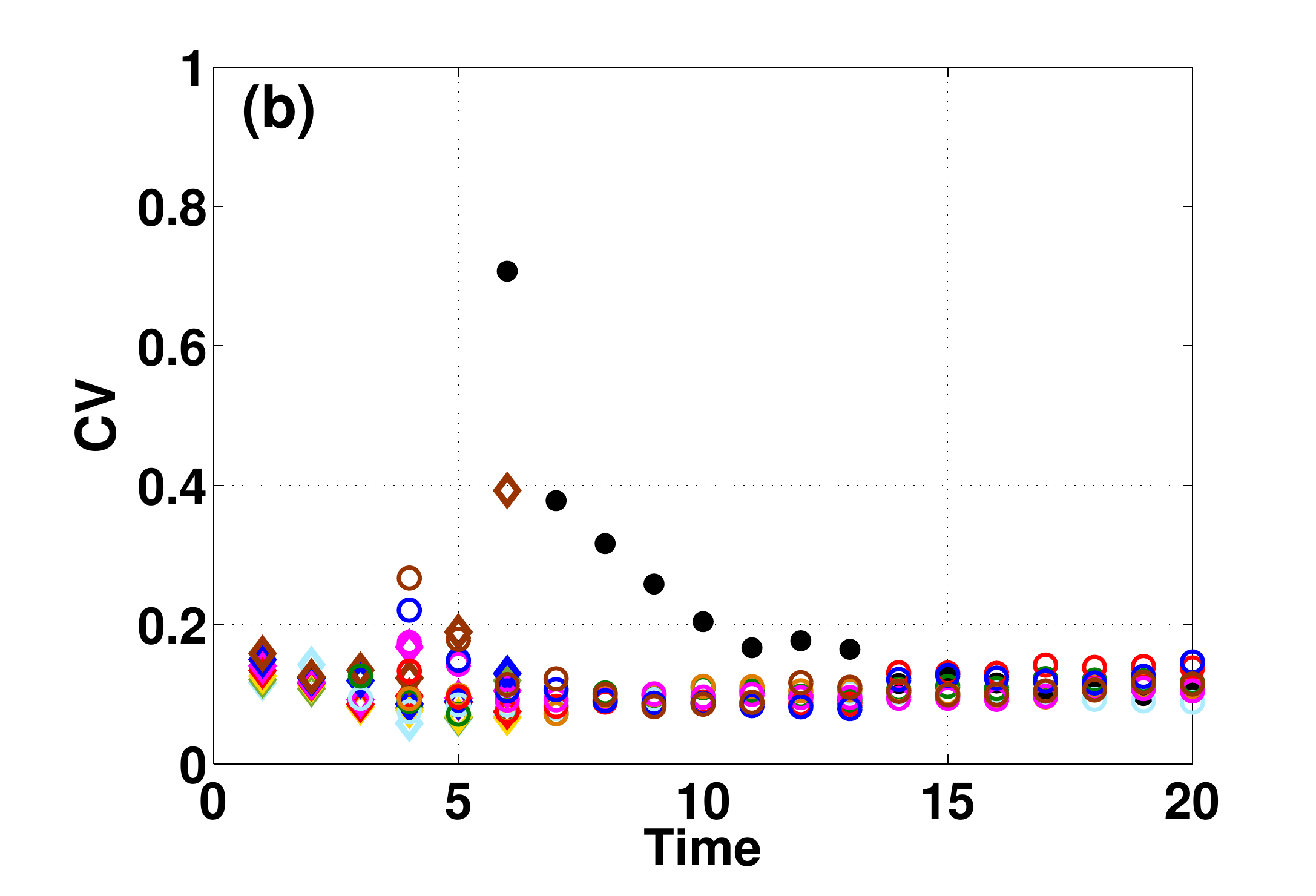}}
          \caption{IS estimate for switching probability (RSER) and CV
            of the RSER with sample size $M = 10^3$ for a weakly
            coupled two-spin system. (a) RSER with applied currents
            ranging from $0$ to $0.6$ versus time $T$ with thermal
            stability factor $\Delta = 60$ and coupling strength
            $ c = 0.2$. (b) CVs of IS estimates in (a). In both
            panels, open circles and open diamonds denote estimates
            generated by IS using infinite-time and finite-time bias
            functions, respectively, and black dots denote estimates
            generated by naive MC simulations for $I_J = 0.6$ with
            sample size $M = 10^5$. The colors correspond to different
            current amplitudes indicated in (a). }
      \label{figs:ISvsMC2_short}
\end{figure*}

\begin{figure*}[ht]
   \centering
          \subfigure{\includegraphics[width =3.4in]{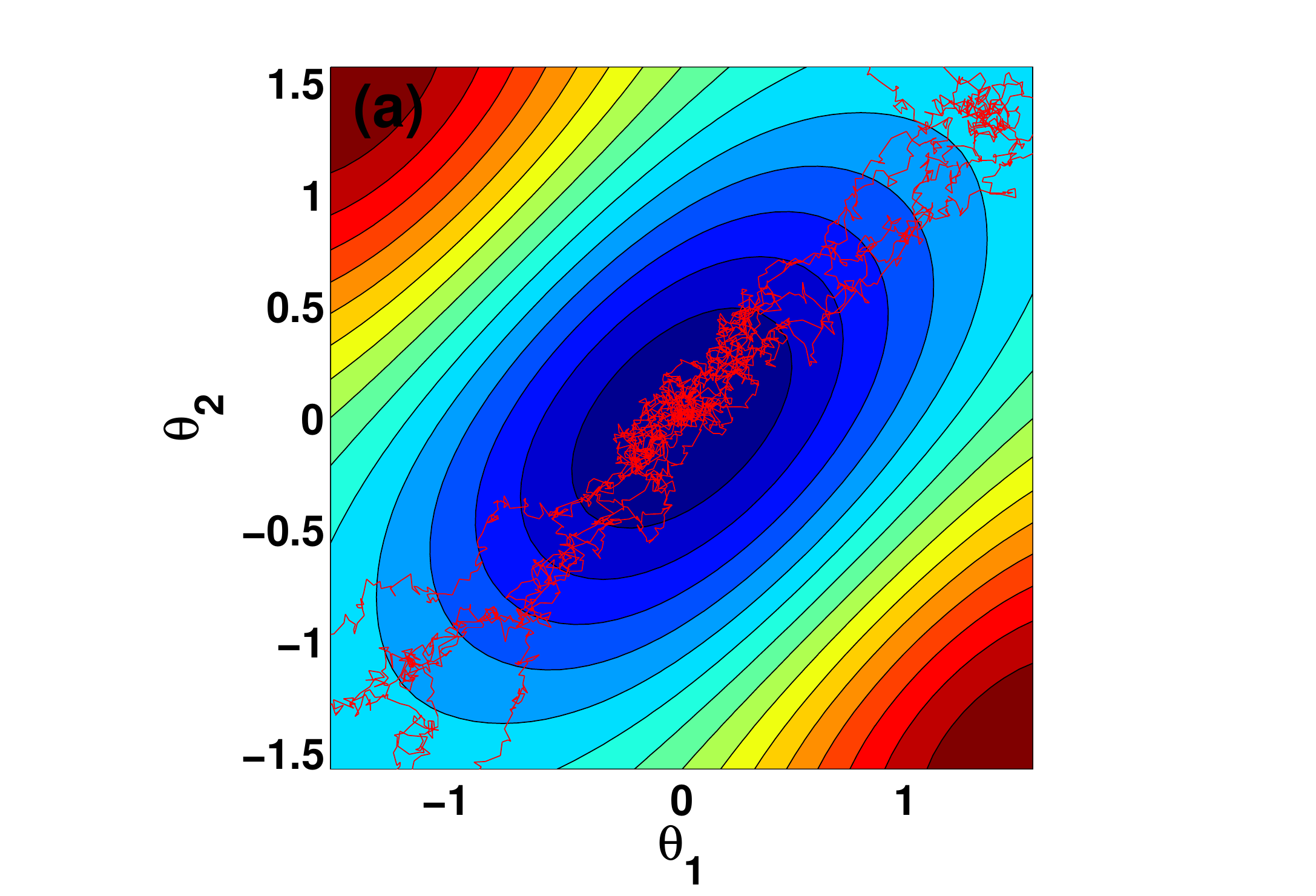}}
          \subfigure{\includegraphics[width =3.4in]{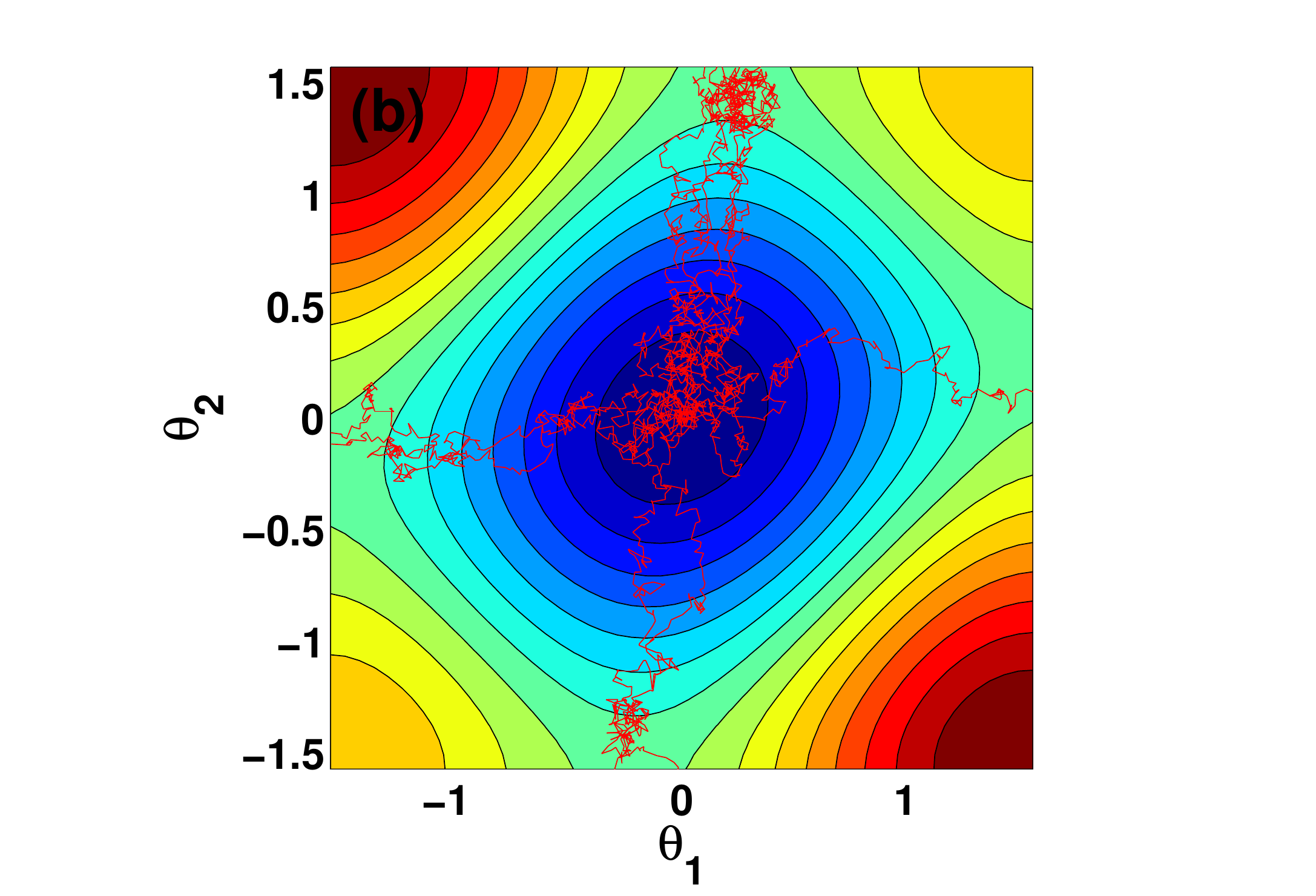}}
          \caption{Sample paths from IS for two-spin systems with
            current $I_J = 0.1$ and thermal stability factor
            $\Delta = 60$: strongly coupled spins with $c = 0.8$ (a),
            weakly coupled spins with $c = 0.2$ (b).  False color
            corresponds to the effective potential in~(\ref{VN})}
      \label{figs:Sample_path}
\end{figure*}

Fig. \ref{figs:ISvsMC1_short} (a) shows IS estimates of switching
probabilities obtained using finite-time bias functions and
infinite-time bias functions with a suitable cutoff near the origin
applied to the coupled system with non-dimensional temperature
$\Delta = 60$ and coupling strength $ c = 0.8$. The coupling strength
$c = 0.8$ is an example of strong coupling, for which the spins rotate
coherently along optimal switching paths. The open circles and open
diamonds denote estimates generated by IS using infinite-time and
finite-time bias functions, respectively. With sampling size of
$M=10^3$, the infinite-time bias functions allow us to sample
switching probabilities for $T \ge 5$ with $I_J$ ranging from $0$ to
$0.6$.~For $T \le 4$, finite-time bias functions are used with IS. For
infinite-time biasing the bias is switched off, i.e., $u^*_\infty$ is
set to zero, when the effective potential from \eqref{VN} of a
configuration $(\tilde \theta_1, \tilde \theta_2)$ falls below that of
$(\theta_0, \theta_0)$ with $\theta_0 \in [0, 0.3]$ chosen to minimize
the CV.

Fig.~\ref{figs:ISvsMC1_short} (b) shows the CVs becoming larger as
time $T$ decreases, indicating that the infinite-time bias functions
become less efficient.  The effectiveness of using finite-time bias
functions is evident in the CV values in Fig.~\ref{figs:ISvsMC1_short}
(b), where the CV values with open diamonds at time $T = 4$ are much
smaller than the CV values with open circles at times $ T =4$ and
$T = 5$.  This improved effectiveness comes with a cost of computing
updated finite-time bias functions at each time step.

Similar results for a weakly coupled system are shown in
Fig.~\ref{figs:ISvsMC2_short}.  Fig.~\ref{figs:ISvsMC2_short}~(a)
shows the switching probabilities as a function of time for different
read currents with non-dimensional temperature $\Delta = 60$ and
coupling strength $ c = 0.2$. The coupling strength $c = 0.2$ is an
example of weak coupling, for which the spins rotate incoherently
along optimal switching paths.  With a sampling size of $M=10^3$, IS
using infinite-time bias functions allows us to sample switching
probabilities for $T \ge 4$ with $I_J$ ranging from $0$ to $0.6$.  As
in Fig.~\ref{figs:ISvsMC1_short}, Fig.~\ref{figs:ISvsMC2_short}~(b)
shows a decrease in efficiency of IS using infinite-time bias
functions as $T$ decreases.

Fig.~\ref{figs:ISvsMC2_short} (a) also shows switching probabilities
generated using naive MC simulations with a sample size of $M = 10^5$
for read current $I_J=0.6$ at various pulse durations.  Naive MC
simulations at this sample size fail to accurately capture
probabilities less than $10^{-5}$ while IS is able to estimate
probabilities as low as $10^{-28}$. This is reflected in
Fig.~\ref{figs:ISvsMC2_short}~(b) where the CV is seen to diverge as
the probability estimate decreases.  Finally, a few representative
switching trajectories corresponding to the results in
Figs. \ref{figs:ISvsMC1_short} and \ref{figs:ISvsMC2_short} are shown
in Fig. \ref{figs:Sample_path}.

  \begin{figure}[ht]
   \centering
   \includegraphics[width
   =3.4in]{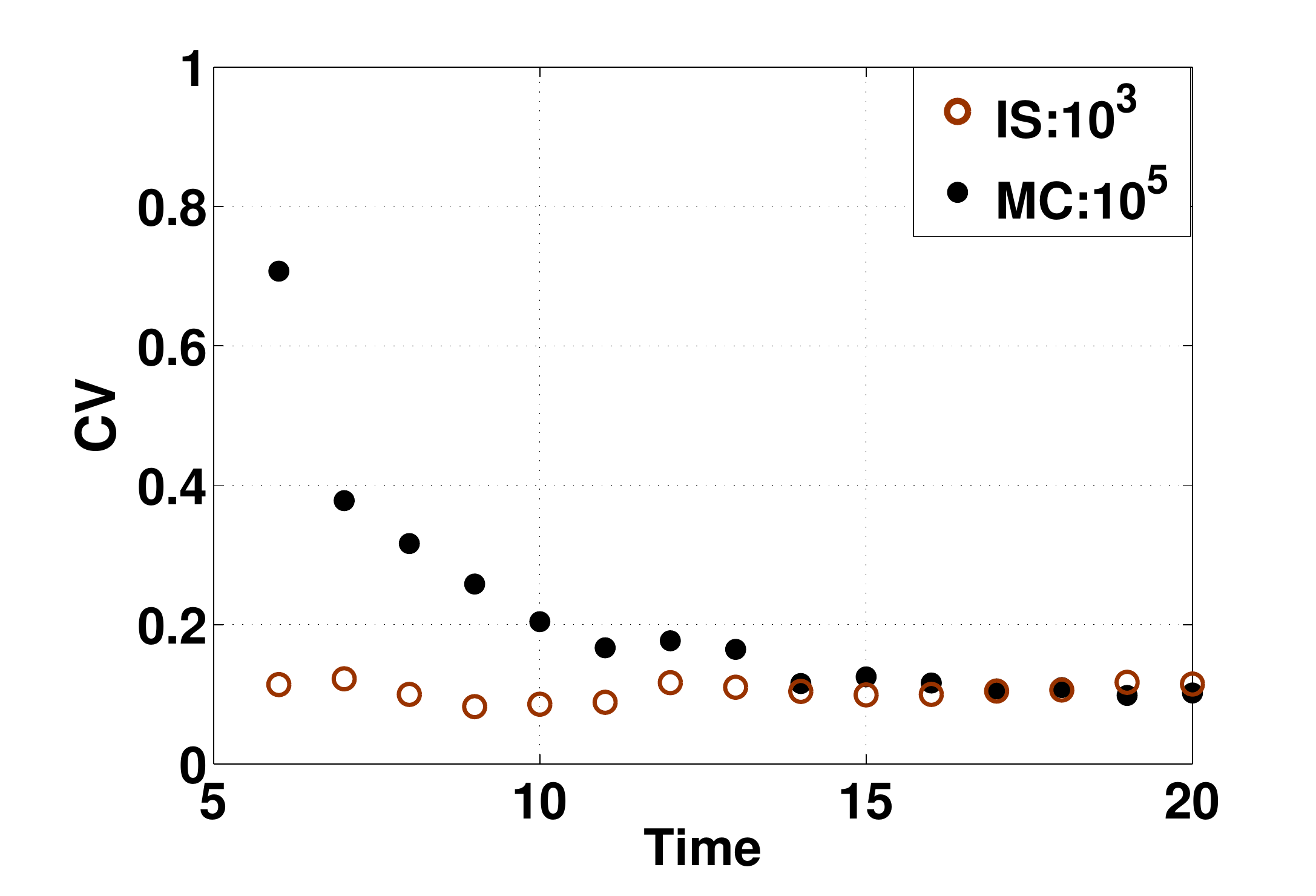}
   \caption{CVs of estimates generated by IS (brown open circles) and
     naive MC (black dots) in Fig. \ref{figs:ISvsMC2_short}. }
   \label{figs:ISvsMC1_2}
\end{figure}
Fig. \ref{figs:ISvsMC1_2} shows a comparison between the CV for IS
with sample size of $10^3$ and the CV for naive MC with sample size of
$10^5$ for $I_J = 0.6$. For time $6 \leq T \leq 15$ the CVs for naive
MC exceed the CVs for IS, by a factor ranging from 1 to 6 as the value
of $T$ is decreased. This means that the number of IS samples required
to achieve an estimate with the same or better accuracy than an
estimate generated by naive MC is smaller by several orders of
magnitude. For example, for switching probability by time $T = 6$,
where the CV ratio is about $6.2$, al least $3.8 \times 10^6$ naive MC
samples are required to achieve the same accuracy of an IS estimate
generated using only $M=10^3$. This contrast is even more stark for
smaller applied currents with much lower associated switching
probabilities, where the computational effort required by naive MC
simulations is prohibitive.  Sampling-based probability estimates are
only available using IS at these parameter values.

\subsection{Failure to switch (WSER)}
In this section, we estimate the WSER using MC and IS for both the
macrospin model and two coupled-spin system, where the current is set
sufficiently high as to drive the origin unstable, such that a
switching event is expected to occur in the majority of random
realizations. The two-spin state is considered to have switched when
$\max(|\theta_1(T_{sw})|, |\theta_2(T_{sw})|) = \pi/2$ for some
$0 < T_{sw} \leq T.$

 \begin{figure}[h]
   \centering
         \includegraphics[width =3.5in]{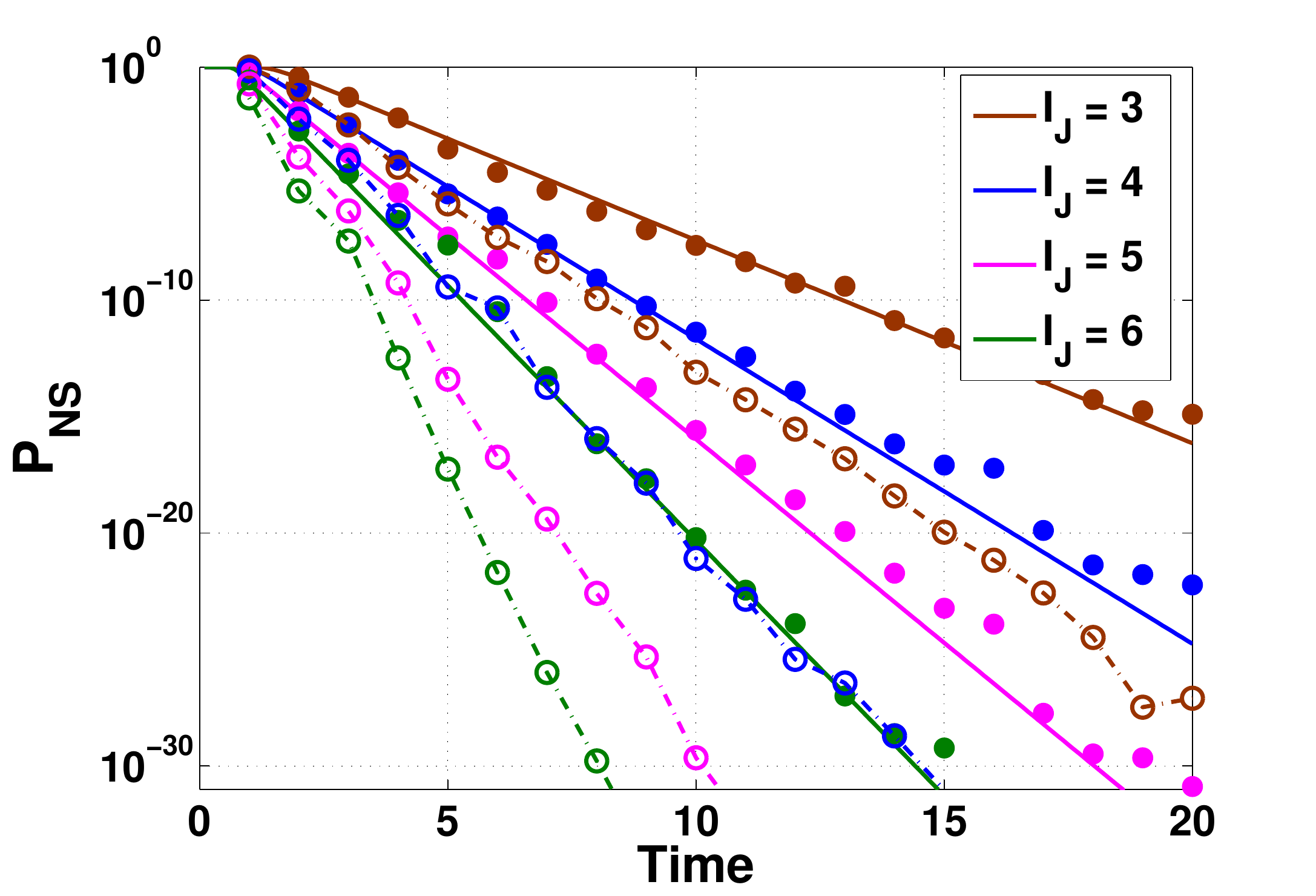}
         \caption{Non-switching probability as a function of time
           horizon for several values of the writing current, using
           thermal stability factor $\Delta = 60$. The current $I_J$
           is identified by color and ranges from 3 to 6. Solid lines
           denote numerical solutions of the backward Fokker-Planck
           equation for single macrospin, while dashed lines are
           simply visual guides.  Filled circles denote IS estimates
           of the non-switching probabilities for a single
           macrospin. Open circles denote IS estimates of
           non-switching probabilities for two coupled spins with
           $c = 0.2$.}
   \label{fig:NS_maroc}
\end{figure}
Fig. \ref{fig:NS_maroc} shows a comparison between the numerical
solution of the FPE and simulation results using IS for a single
macrospin. Independent realizations of the macrospin evolution were
computed with the same initial conditions, currents $I_J=3,4,5$, and
$6$, and independent thermal noise with $\Delta=60$. The estimates
obtained using $M=10^4$ IS samples show good agreement with the
numerical solution of the backward FPE for moderate and long
times. For the two-spin system, $10^3$ IS samples are used to estimate
non-switching probabilities.
\medskip

\section{Summary}
\label{s:discussion}

In conclusion, we have applied IS to estimate error rates in reading
and writing spin-torque memory devices in the macrospin limit and for
the case of two coupled spins, using a variety of applied currents and
thermal stability factors.  We have demonstrated how IS is able to
compute probabilities well below those computable using naive MC
simulations, while producing accurate estimates with improved
efficiency in cases where the probability is computable using naive MC
but still very small.  Depending on the time horizon of the read or
write event relative to the drift dynamics of the spin system, IS
simulations using infinite- or finite-time bias functions are
appropriate.  Infinite-time bias functions require only moderate
computational cost; however, the increased cost can be significant in
computing finite-time bias functions. Further improvement of the
infinite-time biasing can be achieved by introducing a threshold turn
the bias on only some distance away from the metastable equilibrium
whose thermal stability is investigated.

\appendix

In this appendix we derive a reduced model for the macrospin dynamics
of the in-plane magnetization component in the regime of
\eqref{eq:Qsmall}, which is done in the spirit of
\cite{kohn05,garcia01,Lund2016}. Our starting point is \eqref{eq:LL}
for a macrospin $\mathbf m$, in which $\mathbf h$ satisfies
\eqref{eq:hmacro}. Introducing cylindrical coordinates, we can write
\begin{equation}
  \mathbf m = \left( \sqrt{1 - z^2} \, \cos \theta , \sqrt{1 -
  z^2} \, \sin \theta, z \right).
\end{equation}
Then, changing the unit of time to $\tau_Q$ we arrive, after some
algebra and a change from Stratonovich to It\^o formulation, to the
following It\^o SDEs for $\theta(t)$ and $z(t)$:
\begin{align}
  \label{sys:polar3}
  {Q (1 + \alpha^2) \over \alpha} \dot{\theta}
  & = -\frac{\partial
    E}{\partial z} - \frac{\alpha}{1-z^2} \frac{\partial E}{\partial
    \theta} +(a_J + \alpha b_J) \frac{\sin \theta}{\sqrt{1-z^2}}
    \nonumber \\
  & + (b_J -
    \alpha a_J) \frac{z \cos \theta}{\sqrt{1-z^2}}
    +Q \sqrt{\frac{1 + \alpha^2}{\Delta(1-z^2)}} \, \dot{W_1},
  \\
  {Q (1 + \alpha^2) \over \alpha} \dot{z}
  & = \frac{\partial E}{\partial \theta} - \alpha(1-z^2)
    \frac{\partial
    E}{\partial z} -{Q
    \alpha z \over \Delta} \nonumber \\
  & + (a_J + \alpha b_J)  z {\sqrt{1-z^2}} \, \cos
    \theta   \nonumber \\
  & - (b_J - \alpha a_J)  \sqrt{1-z^2} \, \sin \theta \nonumber \\
  & +
    Q \sqrt{\Delta^{-1} (1 + \alpha^2)(1-z^2)} \, \dot{W_2},
\end{align}
where $W_1$ and $W_2$ are two uncorrelated Brownian motions and
\begin{align}
  \frac{\partial E}{\partial z}  =  z - Q z \sin^2 \theta, \quad
  \frac{\partial E}{\partial \theta}  = Q (1-z^2) \, \sin \theta \cos
                                        \theta .
\end{align}

We now assume that $Q \ll 1$, while $a_J = \beta^{-1} I_J Q$,
$b_J = I_J Q$ and $\epsilon = Q / (2 \Delta)$, with $I_J$, $\beta$ and
$\Delta$ of order unity. In this case any deviations of $z$ from zero
are strongly suppressed. Therefore, linearizing the above equations in
$z$ and introducing $\zeta = Q^{-1} z$, we obtain
\begin{align}
  \label{sys:polar3lin}
  {Q (1 + \alpha^2) \over \alpha} \dot{\theta}
  & = -\frac{\partial
    E}{\partial z} - \alpha \frac{\partial E}{\partial
    \theta} + Q I_J (\beta^{-1} + \alpha) \sin \theta
    \nonumber \\
  &
    + Q^2 I_J (1 -
    \alpha \beta^{-1}) \zeta \cos \theta \nonumber \\
  & +Q \sqrt{\Delta^{-1} (1 + \alpha^2)} \, \dot{W_1},
  \\
  {Q^2 (1 + \alpha^2) \over \alpha} \dot{\zeta}
  & = \frac{\partial E}{\partial \theta} - \alpha \frac{\partial
    E}{\partial z} -{Q^2 \alpha \zeta \over \Delta} \nonumber \\
  & + Q^2 I_J  (\beta^{-1} + \alpha )  \zeta \cos
  \theta   \nonumber \\
  & - Q I_J  (1 - \alpha
    \beta^{-1})  \sin \theta \nonumber \\
  & +
    Q \sqrt{\Delta^{-1} (1 + \alpha^2)} \, \dot{W_2},   \label{sys:polar3linn}
\end{align}
where
\begin{align}
  \label{dE}
  \frac{\partial E}{\partial z}  =  Q \zeta , \qquad
  \frac{\partial E}{\partial \theta}  = Q \sin \theta \cos
  \theta .
\end{align}
Substituting \eqref{dE} into \eqref{sys:polar3lin} and
\eqref{sys:polar3linn}, to the leading order in $Q \ll 1$ we then
arrive at
\begin{align}
  {(1 + \alpha^2) \over \alpha} \dot{\theta}
  & = - \zeta - \alpha \sin \theta \cos \theta + I_J (\beta^{-1} +
    \alpha) \sin \theta
    \nonumber \\
  &
    +\sqrt{\Delta^{-1} (1 + \alpha^2)} \, \dot{W_1}, \label{theta}
  \\
  0
  & =  \sin \theta \cos \theta - \alpha \zeta -  I_J  (1 - \alpha
    \beta^{-1})  \sin \theta \nonumber \\
  & +
    \sqrt{\Delta^{-1} (1 + \alpha^2)} \, \dot{W_2}, \label{zeta}
\end{align}
Finally, solving for $\zeta$ in \eqref{zeta} and substituting it back
into \eqref{theta}, with the help of the fact that
$\alpha W_1 - W_2 = \sqrt{1 + \alpha^2} \, W$, where $W$ is another
Brownian motion, we obtain \eqref{sys:macro}. It is interesting to
note that only the contribution of the field-like spin torque appears
in the reduced equation, while the contribution of the Slonczewski
torque cancels out to the leading order.

%

%
%

\section*{Acknowledgments}

The work of YMY and CBM was supported, in part, by NSF via grants
DMS-1313687 and DMS-1614948. YMY would like to thank the Department of
Mathematics at the University of Illinois at Urbana-Champaign, where
part of this work was done, for its hospitality. CBM wishes to
acknowledge valuable discussions with E. Vanden-Eijnden.

\ifCLASSOPTIONcaptionsoff
  \newpage
\fi



\bibliographystyle{IEEEtran}
\bibliography{citation}

\end{document}